\documentclass[useAMS,usenatbib]{mn2e}
\usepackage{amsmath}
\usepackage{txfonts}
\usepackage{natbib}
\usepackage[all]{xy}
\usepackage{graphicx}

\def\del#1{{}}

\newcommand{\ltsima}{$\; \buildrel < \over \sim \;$}
\newcommand{\lsim}{\lower.5ex\hbox{\ltsima}}
\newcommand{\gtsima}{$\; \buildrel > \over \sim \;$}
\newcommand{\gsim}{\lower.5ex\hbox{\gtsima}}
\newcommand{\bra}{\langle}
\newcommand{\ket}{\rangle}

\newcommand{\dd}{\mathrm{d}}
\newcommand{\e}{\mathrm{e}}

\newcommand{\dirac}{\delta_\mathrm{D}}

\newcommand{\be}{\begin{equation}}
\newcommand{\ee}{\end{equation}}
\newcommand{\bml}{\begin{multline}}
\newcommand{\eml}{\end{multline}}
\newcommand{\ba}{\begin{eqnarray}}
\newcommand{\ea}{\end{eqnarray}}
\newcommand{\nn}{\nonumber}
\newcommand{\mb}{\mathbf}
\newcommand{\ph}{\varphi} 
\newcommand{\bk}{\mb k}
\newcommand{\bl}{\bmath \ell}
\newcommand{\bn}{\mb n}
\newcommand{\bq}{\mb q}
\newcommand{\bQ}{\mb Q}

\newcommand{\br}{\mb r}
\newcommand{\bx}{\mb x}

\title[Cross iSW-galaxy bispectra and trispectra]
{Cross bispectra and trispectra of the non-linear integrated Sachs-Wolfe effect and the tracer galaxy density field}
\author[G. J{\"u}rgens and B.M. Sch{\"a}fer]{
Gero J{\"u}rgens\thanks{gero.juergens@stud.uni-heidelberg.de}$^1$ and Bj{\"o}rn Malte Sch{\"a}fer$^2$\\
$^1$Institut f{\"u}r theoretische Astrophysik, Zentrum f{\"u}r Astronomie, Universit{\"a}t Heidelberg, Philosophenweg 12, 69120 Heidelberg, 
Germany\\
$^2$Astronomisches Recheninstitut, Zentrum f{\"u}r Astronomie, Universit{\"a}t Heidelberg, M{\"o}nchhofstra{\ss}e 12, 69120 Heidelberg, Germany}

\begin{document}
\pagerange{\pageref{firstpage}--\pageref{lastpage}}
\pubyear{2012}
\maketitle
\label{firstpage}

\begin{abstract}
In order to investigate possibilities to measure non-Gaussian signatures of the non-linear iSW effect, we study in this work the family of mixed bispectra 
$\bra\tau^{\,q}\gamma^{\,3-q}\ket$ and trispectra $\bra\tau^{\,q}\gamma^{\,4-q}\ket$ between the integrated Sachs-Wolfe (iSW) temperature perturbation $\tau$ 
and the galaxy over-density $\gamma$. We use standard Eulerian perturbation theory restricted to tree level expansion for predicting the cosmic matter field. 
As expected, the spectra are found to decrease in amplitude with increasing $q$. The transition scale between linear domination 
and the scales, on which non-linearities take over, moves to larger scales with increasing number of included iSW source 
fields $q$. We derive the cumulative signal-to-noise ratios for a combination of \textit{Planck} CMB data and the galaxy sample of a \textit{Euclid}-like survey. 
Including scales down to $\ell_\mathrm{max}=1000$ we find sobering values of $\sigma \simeq 0.83$ 
for the mixed bispectrum and $\sigma \simeq0.19$ in case of the trispectrum for $q=1$. For higher values of $q$ the polyspectra $\bra\tau^2\gamma\ket$ and $\bra\tau^3\gamma\ket$ 
are found to be far below the detection limit.
\end{abstract}


\section{INTRODUCTION}
The integrated Sachs-Wolfe (iSW) effect is one of the secondary anisotropies of the cosmic microwave background (CMB). Time-evolving gravitational 
potentials in the large-scale structure generate temperature fluctuations in the CMB \citep{Sachs1967}. The linear part of this effect is a valuable 
tool for investigating dark energy and non-standard cosmologies since it is sensitive to fluids with non-zero equation of state \citep{Crittenden1996}. 
For this reason its detection is of particular relevance for cosmology and the nature of gravity \citep{Lue2004,Zhang2006} even though its signal strength is very low.

The linear iSW effect has been measured in such cross-correlation studies \citep{Boughn1998,Boughn2004,Vielva2006,mcewen2007,Giannantonio2008}. 
There are, however, doubts on detection claims formulated by
\citet{Hern'andez-Monteagudo2010} and \citet{L'opez-Corredoira2010}, who point out that the iSW-signal seems to be too weak on low multipoles below $\ell \sim 10$,
and that field-to-field fluctuations and sampling errors can be important. These facts may correct the detection significance to-date to a number of less than two.

While the linear iSW signal is a large scale effect and becomes negligible at angular wave numbers above $\ell\sim100$, non-linear evolution 
of the gravitational potential and leaves signatures on much smaller scales, also called Rees-Sciama effect \citep{Rees1968} 
and surpasses the linear iSW-effect on these scales. 
The possible signatures of this effect in angular cross spectrum have been thoroughly studied analytically \citep{Martinez-Gonzalez1994,
Sanz1996, Seljak1996, Schafer2006}. The effect increases the total iSW signal by roughly two orders of magnitude at angular scales around $\ell \sim 1000$ \citep{Cooray2002}, 
before gravitational lensing and kinetic Sunyaev-Zel'dovich effect become dominant at even smaller scales. 
However, comparisons of theoretical studies with numerical simulations showed the Rees-Sciama to be negligible in comparison with primary 
anisotropies on angular scales larger than $\theta > 1^\prime$ \citep{Tuluie1995, Seljak1996}. Also from cross-correlations of the 
CMB with weak lensing surveys only a detection significance of $\sim 1.5 \sigma$ from $Planck+$LSST is expected \citep{Nishizawa2008}. 


One option to obtain direct signatures of non-Gaussianities is the investigation of higher order connected correlators \citep{Schafer2008}. 
In this work we aim to formulate a perturbative approach of the mixed iSW-galaxy polyspectra, concentrating on the tree-level 
bispectra and trispectra in flat sky approximation. The unequal 
rate of linear and non-linear evolution at different scales will lead to interesting sign changes in the spectra, which will also be apparent in the 
non-trivial time evolution of the different source field contributions. In addition, we will study the signal-to-noise spectra for measurements expected from 
\textit{Planck} CMB data in cross-correlation with observations from a \textit{Euclid}-like survey assuming unbiased measurements with Gaussian noise contributions. 
We revisit a previous estimate of the observability of the iSW-bispectrum \citep{Schafer2008} correcting an error in the expression 
for the spectrum of the gravitational potential and because of the significantly improved signal-to-noise computation, which uses an 
adaptive Monte-Carlo integration scheme \citep{Hahn2005} instead of a binned summation over the multipoles.

The article has the following structure: In Section~\ref{sect_founds} we will lay out the theoretical framework for linear and non-linear structure formation 
(Section~\ref{sect_lsf} and Section~\ref{sect_nlsf}), as well as for the theory of higher order correlators of the density field in Section~\ref{sect_pt-trispectrum}.
Furthermore, the main fields of interest, the galaxy number distribution (Section~\ref{sect_galaxy_distribution}) and the iSW-effect 
(Section~\ref{sect_galaxy_distribution}) are introduced.
The mixed bispectra and trispectra are discussed in Section~\ref{sect_mixed}, with specific studies of their weighting functions (Section~\ref{sect_weighting}) 
and their time evolution (Section~\ref{sect_time_evo}). 
In Section~\ref{sect_detect} we present the relevant noise sources (Section~\ref{sect_noise}), the resulting covariances (Section~\ref{sect_cov}) of the polyspectra 
and finally derive their signal-to-noise ratios (Section~\ref{sect_s2n}). 
Our results are summarized and discussed in Section~\ref{sect_sum}.

The reference cosmological model used is a spatially flat $\Lambda$CDM cosmology with Gaussian adiabatic initial perturbations 
in the cold dark matter density field. The specific parameter choices are 
$\Omega_{\mathrm m} = 0.25$, $n_{\mathrm s} = 1$, $\sigma_8 = 0.8$, $\Omega_\mathrm{b}=0.04$ and $H_0=100\: h\:\mathrm{km}/\mathrm{s}/\mathrm{Mpc}$, with $h=0.72$. 
\section{FOUNDATIONS}\label{sect_founds}

\subsection{Dark energy cosmologies}\label{sect_de_cosmo}
In spatially flat dark energy cosmologies with the matter density parameter $\Omega_{\mathrm m}$, the Hubble function $H(a)=\dd\ln a/\dd t$ is given by
\begin{equation}
\frac{H^{\,2}(a)}{H_0^{\,2}} = \Omega_{\mathrm m}\,a^{\,-3} + (1-\Omega_{\mathrm m})\,a^{\,-3\,(1+w)},
\end{equation}
with a constant dark energy equation of state parameter $w$. The value $w\equiv -1$ corresponds to the cosmological constant $\Lambda$. 
The relation between comoving distance $\chi$ and scale factor $a$ is given by
\begin{equation}
\chi = c\int_a^1\dd a\:\frac{1}{a^2 H(a)},
\end{equation}
in units of the Hubble distance $\chi_H=c/H_0$.

\subsection{CDM power spectrum}\label{sect_cdm_ps}
The CDM density power spectrum $P(k)$ describes the fluctuation amplitude of the Gaussian homogeneous density field 
$\delta(\bk)$, $\bra\delta(\bk)\delta^*(\bk^\prime)\ket=(2\pi)^3\dirac(\bk-\bk^\prime)P(k)$, and the power spectrum of linear evolving 
density fields $\delta_\mathrm{L}(\bk)$ is given by the ansatz
\begin{equation}
P_\mathrm{L}(k)\propto k^{\,n_{\mathrm s}}T^2(k),
\end{equation}
with the transfer function $T(k)$. In low-$\Omega_{\mathrm m}$ cosmologies $T(k)$ is approximated with the fit proposed by \cite{Bardeen1986},
\begin{eqnarray}
T(q) &= &\frac{\ln(1+2.34\,q)}{2.34\,q}\nonumber \\
   & &\times \,\left[1+3.89\,q+(16.1\,q)^2+(5.46\,q)^3+(6.71\,q)^4\right]^{-1/4}\hspace{0.1 cm},
\label{eqn_cdm_transfer}
\end{eqnarray}
where the wave number $k=q\,\Gamma$ is rescaled with the shape parameter $\Gamma$ \citep{Sugiyama1995} 
which assumes corrections due to the baryon density $\Omega_\mathrm{b}$,
\begin{equation}
\Gamma=\Omega_{\mathrm m} h\,\exp\left[-\Omega_\mathrm{b}\left(1+\frac{\sqrt{2h}}{\Omega_{\mathrm m}}\right)\right].
\end{equation}
The spectrum $P(k)$ is normalized to the variance $\sigma_8$ on the scale $R=8~\mathrm{Mpc}/h$,
\begin{equation}
\sigma^2_R 
= \frac{1}{2\pi^2}\int\dd k\:k^2 P(k) W^2(kR)
\end{equation}
with a Fourier transformed spherical top hat filter function, $W(x)=3j_1(x)/x$. $j_\ell(x)$ is the 
spherical Bessel function of the first kind of order $\ell$  \citep{Abramowitz1972}. 
\subsection{Linear structure growth}\label{sect_lsf}
The linear homogeneous growth of the relative density perturbation field $\delta_\mathrm L(\bmath x, a)$ is described by the growth function $D_+(a)$
\be
\delta_\mathrm L(\bk,a)=D_+(a)\,\delta_\mathrm L(\bk,a=1)\,,
\ee 
which is the solution to the growth equation \citep{Turner1997,Wang1998,Linder2003},
\begin{equation}
\frac{\dd^2}{\dd a^2}D_+(a) + \frac{1}{a}\left(3+\frac{\dd\ln H}{\dd\ln a}\right)\frac{\dd}{\dd a}D_+(a) = 
\frac{3}{2\,a^2}\,\Omega_{\mathrm m}(a) \,D_+(a) \hspace{0.1 cm}.
\label{eqn_growth}
\end{equation}
The growth equation can be obtained by combining the linearized structure formation equations consisting of Poisson equation, 
Euler equation and continuity equation. 

Considering the special case of flat SCDM cosmologies, where $\Omega_\mathrm m \equiv 1$, the Hubble function scales like $H=H_0\,a^{-3/2}$. The growing mode solution 
then gives the very simple growth function $D_+(a)=a$, which is even in more complex cosmologies a good approximation during the matter domination era.
\subsection{Non-linear structure formation}\label{sect_nlsf}
In order to describe non-Gaussianities in the density source field generated by non-linear evolution a theoretical approximation is required. 

We employ non-linear solutions to the density field from standard Eulerian perturbation theory \citep{Sahni1995,Bernardeau2002}. 
One expands the density contrast $\delta(\bk, a)$ in $n$-th order perturbative contributions $\delta^{(n)}(\bk, a)$, 
which can be written in terms of the perturbation theory kernels $F^{(n)}(\bk_1,...,\bk_n)$ and the initial linear fields 
$\delta^{(1)}(\bk) = \delta_{\,\mathrm L}(\bk, a=1)$:
\ba
 \delta (\bk, a) &=& \sum_{n=1}^{\infty} \, D_+^n \,\delta^{(n)}(\bk) \label{pt-expansion}\\
 \delta^{(n)}(\bk) &=& \int\mathrm d^3 q_1\, ...\int\mathrm d^3 q_n\, \dirac(\bk-\bq_{1...\,n}) \nn \\
~&~&\times \,F^{(n)}(\bq_1,...,\bq_n)\, 
\delta^{(1)}(\bq_1)\,...\,\delta^{(1)}(\bq_n)\label{pt-modes}
\ea
with $\mb q_{1...\,n}=\mb q_1+ ... +\mb q_{\,n}$. 
By inserting eqs.~(\ref{pt-expansion}-\ref{pt-modes}) into the evolution equations one finds recursion relations for the kernels $F^{(n)}(\bq_1,...,\bq_n)$ 
by combinatorics \citep{Goroff1986,Jain1994}.
The explicit symmetrized expressions for the second order perturbation theory kernels take a very simple and intuitive form:
\ba
F^{(2)}(\bk_1,\bk_2) &=& \frac 57 + \frac 12 \frac{\bk_1\cdot \bk_2}{k_1 k_2}\left( \frac{k_1}{k_2} + \frac{k_2}{k_1}\right) 
+ \frac27 \frac{(\bk_1\cdot \bk_2)^2}{k_1^2 k_2^2}\\
G^{(2)}(\bk_1,\bk_2) &=& \frac 37 + \frac 12 \frac{\bk_1\cdot \bk_2}{k_1 k_2}\left( \frac{k_1}{k_2} + \frac{k_2}{k_1}\right) 
+ \frac47 \frac{(\bk_1\cdot \bk_2)^2}{k_1^2 k_2^2}\hspace{0.1 cm}\hspace{0.1 cm},
\ea
where $F^{(1)}=G^{(1)} = 1$.
One can see that mode-coupling to second order reaches its maximum when the contributing modes $\bk_1$ and  $\bk_2$ are aligned, 
whereas the kernel vanishes for anti-parallel modes. When in eqn.~(\ref{pt-modes}) $n$ different modes $\mb q_1...\mb q_n$ contribute to a mode $\mb k$, wave number 
conservation holds, enforced by the $\dirac$-distribution: $\mb k = \mb q_1+...+\mb q_n$.

\subsection{The $n$-point functions in perturbation theory}\label{sect_pt-trispectrum}
For an analytic expression of the perturbation theory $n$-point function one has to expand the fields in the correlator. Due to the assumed Gaussianity of the 
initial field $\delta^{(1)}$ the correlators with an even number of fields $\delta^{(1)}$ will later simplify to products of initial two-point functions 
$P_\mathrm L(k)$, while all uneven contributions vanish: 
\be
\bra \delta_1...\delta_n \ket = \left\bra \sum_{i_1} D_+^{i_1}\,\delta_1^{(i_1)}\,\dots\,\sum_{i_n} D_+^{i_n}\,\delta_1^{(i_n)}\right\ket \hspace{0.1 cm}.
\ee
For simplicity we use in this subsection the notation $\delta_n\equiv\delta (\bk_n)$. 
Simple truncation of the expansion in eqn.~(\ref{pt-expansion}) would lead to an inconsistent inclusion of powers of the linear power spectrum 
$P_\mathrm L(k)$. It is more sensitive to take into account all terms up to a certain power $m$ in the linear power spectrum, 
which is equivalent to including terms with initial fields up to powers $2m$. 

In this work we exclusively use tree-level perturbation theory, i.e. no perturbative terms with wave number integrations are taken into account. 
Following this path, the density bispectrum $B_\delta^{\,\bk_1,\bk_2,\bk_3}$ can be written as
\be
B_\delta^{\,\bk_1,\bk_2,\bk_3} = 2\,F^{(2)}(\bk_1,\bk_2)\,P_\mathrm L (k_1)\,P_\mathrm L (k_2) + \mathrm{cycl.}\,\{1,\,2,\,3\}\,.
\ee
The non-Gaussian part of the 4-point function is the trispectrum $T_\delta^{\,\bk_1,\bk_2,\bk_3,\bk_4}$. It is  convenient to split its 
tree-level expression up into two parts. 
The first contribution originates from second order perturbation theory. In this case, two of the fields in the correlator have been 
expanded to second order. The expressions in terms of the initial power spectra and the second order kernels are of the type
\ba
\label{t22}
t^{\,(2)} ((\bk_1,\bk_2),(\bk_3,\bk_4))&=&  4\,D_+^6\,P_\mathrm{L}(k_3) P_\mathrm{L}(k_4)\times \\
 ~&~&\left(\,F^{(2)}(\bk_{13},-\bk_3)\, F^{(2)}(\bk_{24},-\bk_4)\, P_\mathrm{L}(k_{13})\right.\nn \\
~&~&\left. + F^{(2)}(\bk_{14},-\bk_4)\, F^{(2)}(\bk_{23},-\bk_3)\, P_\mathrm{L}(k_{14})\,\right)\hspace{0.1 cm}.\nn 
\ea
The second contribution is due to third order perturbation theory. Here, one field is expanded to third order while the other three remain at linear order. 
For this reason only one perturbation kernel appears in the expression for this type of contributions
\be
\label{t3}
t^{\,(3)} (\bk_1,\bk_2,\bk_3,\bk_4) = 6\,D_+^6\, \, F^{(3)}(\bk_1,\bk_2,\bk_3)\, 
P_\mathrm{L}(k_1) P_\mathrm{L}(k_2) P_\mathrm{L}(k_3)\hspace{0.1 cm}.
\ee
With these two functions the connected perturbation theory four-point function up to third order in the linear power spectrum $P_\mathrm{L}(k)$ can be 
expressed by the following two tree-level contributions
\ba
T_\delta^{\,\bk_1,\bk_2,\bk_3,\bk_4} &=& \,\,\,\, \,t^{\,(2)}((\bk_1,\bk_2),(\bk_3,\bk_4)) + \hbox{all pairs}\in \{1,\,2,\,3,\,4\} \nn \\
~&~ &+\, t^{\,(3)} (\bk_1,\bk_2,\bk_3,\bk_4) +\, \mathrm{cycl.}\,\{1,\,2,\,3,\,4\}\,.
\ea
The second order and the third order contributions of $T_\delta$, however, have the same time dependence $D_+^6$. 
We will see, that this is not longer the case for mixed trispectra.
\subsection{Galaxy distribution}\label{sect_galaxy_distribution}
Galaxies form when strong peaks in the density field decouple from the Hubble expansion due to self-gravity. These so called protohalos 
undergo an elliptical collapse \citep{Mo1997,Sheth2001}. 

In contrary to the pressure-less dark matter component the baryons inside a 
dark matter halo can loose energy via radiative cooling and form stars. 
Because of the more complex behavior of baryons, the relation between 
the fractional perturbation $\Delta n/\bra n\ket$ in the mean number 
density of galaxies $\bra n \ket$ and the dark matter over-density $\delta=\Delta \rho/\rho$ is not yet understood. 
In a very simple way, however, the linear relation between the two entities,
\be
 \frac{\Delta n}{\bra n \ket} = b\,\frac{\Delta\rho}{\bra \rho \ket}\hspace{0.1 cm},
\ee
is a good approximation in most cases and was proposed by \cite{Bardeen1986}. 
The bias parameter $b$ can generally depend on a number of variables but 
for simplicity we set the 
galaxy bias to unity throughout this paper, $b\equiv1$.
An established parametrization of the redshift distribution $n(z)\,dz$ of galaxies is
\be
 n(z)\,\dd z = n_0\, \left(\frac z{z_0}\right)^2\, \exp \left[ -\left(\frac z{z_0}\right)^\beta \right]\ \dd z \hspace{0.3 cm}\hbox{with}\hspace{0.3 cm}
 \frac {1}{n_0} = \frac{z_0}{\beta}\, \Gamma\left( \frac 3\beta\right)
\ee
which was introduced by \cite{Smail1995} and will also be used in this work. The parameter $z_0$ is related to the median redshift of the galaxy sample 
$z_{\mathrm{med}}=1.406\, z_0$ if $\beta = 3/2$. For \textit{Euclid} the median redshift is $z_{\mathrm{med}}=0.9$. 
Finally, the $\Gamma$-function \citep{Abramowitz1972} determines the normalization parameter $n_0$.

\subsection{ISW-effect}\label{sect_isw}
Due to its expansion our universe had cooled down sufficiently to allow the formation of hydrogen atoms at a redshift of $z \simeq 1089$ 
\citep{Spergel2003}. Fluctuations in the gravitational potential imposed a shift in the decoupled photons which were emitted in the 
(re)combination process (Sachs-Wolfe effect). This primary anisotropy can be observed in the cosmic microwave background (CMB) 
in form of temperature fluctuations $\Delta T/T_{\mathrm{CMB}} \simeq 10^{-5}$ on large scales around its mean temperature 
$T_{\mathrm{CMB}}=2.726$ K \citep{Fixsen2009}.

Besides this, photons are subjected to several other effects on their way to us, which lead to secondary anisotropies \citep[reviewed by ][]{Aghanim2008}, 
of which only the most important ones are mentioned here: 
Gravitational lensing \citep{Hu2000}, Compton-collisions with free cluster electrons \citep[Sunyaev-Zel´dovich effect,][]{Zeldovich1980} 
and with electrons in uncollapsed structures  \citep[Ostriker-Vishniac effect,][]{Ostriker1986} and gravitational coupling to linear 
\citep[integrated Sachs-Wolfe effect,]{Sachs1967} and non-linear time-evolving potential wells \citep{Rees1968}.
Heuristically, the latter two effects originate from an unbalance between the photon's blue-shift when entering a time varying potential well and the red-shift 
experienced at the exit.

Assuming a completely transparent space, i.e. vanishing optical depth due to Compton scattering, 
the temperature fluctuations $\tau(\hat\theta)$ generated by the iSW-effect can be expressed by the line of sight integral \citep{Sachs1967} 
\be
 \tau(\bmath\theta) \equiv \frac {\Delta T_{\mathrm{iSW}}}{T_{\mathrm{CMB}}} = \frac{2}{c^3} \, \int_0^{\chi_H} \dd\chi\, 
  a^2\, H(a)\,\frac{\partial}{\partial a} \Phi\,(\bmath\theta \chi, \chi)\hspace{0.1 cm} ,
\ee
reaching out to the limit of Newtonian gravity. Using the Poisson equation we can write this integral in terms of the dimensionless 
potential $\phi = \Delta^{-1}\,\delta/ \chi_{H}^2$ of the density field $\delta(\bmath\theta\chi,\chi)$. The $n$-th perturbative order of the 
iSW temperature fluctuation $\tau=\tau^{(1)}+\tau^{(2)}+...$ can 
now be written as
\be
 \tau^{(n)}(\bmath\theta)=\frac{3\,\Omega_\mathrm{m}}{c}\, \int _0^{\chi_H} \dd\chi \, a^2\, H(a) \,\frac{\Delta^{-1}}{\chi_{H}^2} 
	  \left(\frac \dd{\dd a}\,\frac{D_+^{(n)}}a\right) \,  \delta \,(\bmath\theta\chi,\chi)
 \hspace{0.1 cm}.
\ee
The linear effect $(n=1)$ vanishes identically in matter dominated universes $\Omega_\mathrm{m}=1$, since then $D_+/a$ is a constant. Therefore, a non-zero 
iSW-signal will be an indicator of a cosmological fluid with $w\neq0$. After the radiation dominated era it will thus be a valuable tool for investigating 
dark energy cosmologies. The non-linear contributions $(n\geq2)$ are now sourced by time derivatives of the higher perturbative orders of the gravitational potential. 
Therefore, the Rees-Sciama effect is also present in SCDM-cosmology.

In order to identify the sources of the effect it is sensible to investigate the cross correlation  
of the iSW amplitude with the line of sight projected relative galaxy over-density $\gamma = \gamma^{(1)}+\gamma^{(2)}+...$
\be
 \gamma^{(n)}(\bmath \theta) = b \, \int_0^{\chi_H}\dd\chi\, n(z)\, \frac{\dd z}{\dd\chi}\, D_+^{\,n}\, \delta\,(\bmath\theta\,\chi , \chi)\hspace{0.1 cm}.
\ee

Since we are interested in rather small scales, where non-linear effects appear, one can approximate the sphere locally as being plane and perform a Fourier transform
\be
 \gamma(\bmath\ell) = \int \mathrm d ^2 \theta \, \gamma(\bmath\theta)\, \e ^{\displaystyle{- i \,(\bmath{\ell\cdot\theta})}}\hspace{0.1 cm}.
\ee 
The observable $\tau$ can be transformed in analogous way. For later notational convenience we define the two weighting functions 
\ba
 W_{\gamma}(\chi) &=& b\,n(z)\,\frac{\mathrm d z}{\mathrm d \chi} \nn  \\
 W_\tau(\chi) &=& 3\,\Omega_{\mathrm m}\, a^2\, \frac Hc 
\ea
and the time evolution functions to $n$-th order
\ba
Q_{\gamma}^{(n)}(a) &=& D_+^{\,n} \nn  \\
Q_{\tau}^{(n)}(a) &=& \frac {\mathrm d}{\mathrm d a}\left(\frac{D_+^{\,n}}{a}\right)\,.
\ea
\begin{figure}
\includegraphics[width=\columnwidth]{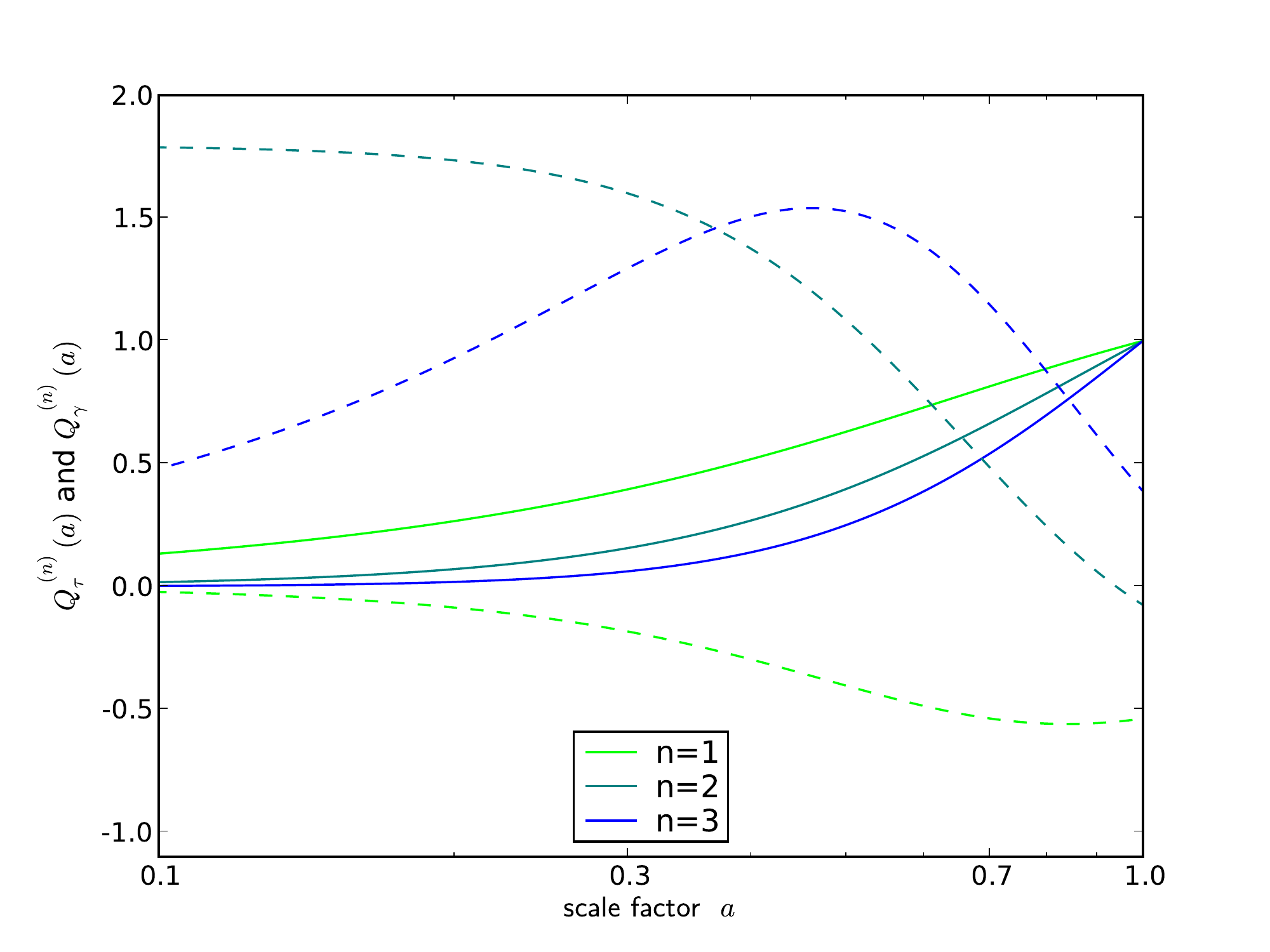}
\caption{Time evolution functions $Q_{\gamma}^{(n)}(a)$ (solid lines) and $Q_{\tau}^{(n)}(a)$ (dashed lines) as a function of the scale factor $a$ for different perturbative orders 
$n=1,2,3$. }
\label{fig_q2_q3}
\end{figure}
In Fig.~\ref{fig_q2_q3} the time evolution functions $Q_{\gamma}^{(n)}(a)$ and $Q_{\tau}^{(n)}(a)$ are depicted in dependence on the scale factor $a$. It is 
particularly interesting to observe the different signs. While the galaxy spectra are always positive, the iSW contributions change their signs with perturbative order. 
As we will later observe, also the signs of $n$-point functions will change consequently in the transition from large scales,
 where the linear theory is valid, to small scales, where non-linearities start to dominate. 
For the cross-bispectrum, this effect has already been studied \citep{Nishizawa2008}.
\section{MIXED BISPECTRA AND TRISPECTRA}\label{sect_mixed}

\subsection{The density polyspectra}\label{sect_multi}
Regardless of the existence of initial non-Gaussianities in the density field $\delta(\bk)$, non-linear structure formation leads 
to non-vanishing three-point and 
higher order correlators due to quadratic terms in the continuity and Euler equation. Since a Gaussian field can uniquely be represented by its two-point 
correlator $\xi (r) = \bra \delta(\bx)\delta(\bx+\br)\ket$, multi-point correlators represent a convenient measure of evolving non-Gaussianities. The Fourier 
transforms of these 2-point and 3-point correlators are the bispectrum $B_\delta^{\,\bk_1, \bk_2, \bk_3}$ and the trispectrum 
$T_\delta^{\,\bk_1, \bk_2, \bk_3,\bk_4}$
\ba
\label{eqn_source_spectra}
\bra \delta (\bk_1)\delta (\bk_2)\delta (\bk_3)\ket &=& (2 \pi)^3 \dirac(\bk_{1...3}) \, B_\delta^{\,\bk_1, \bk_2, \bk_3} \nn \\
\bra \delta (\bk_1)\delta (\bk_2)\delta (\bk_3) \delta(\bk_4)\ket_c &=& (2 \pi)^3 \dirac(\bk_{1...4}) \, T_\delta^{\,\bk_1, \bk_2, \bk_3,\bk_4} \, ,
\ea 
where the Dirac $\dirac$-function is a result of homogeneity.

\subsection{Limber Projection}\label{sect_limber}
In the flat sky approximation one can use a simplified Limber projection \citep{Limber1953} to relate the 3-dimensional 
source spectra $B_\delta^{\,\bk_1, \bk_2, \bk_3}$ and $T_\delta^{\,\bk_1, \bk_2, \bk_3,\bk_4}$ to the angular spectra 
$B_\gamma^{\,\bl_1, \bl_2, \bl_3}$ and $T_\gamma^{\,\bl_1, \bl_2, \bl_3,\bl_4}$.
\ba
B_\gamma^{\,\bl_1, \bl_2, \bl_3} &=& \int_0^{\chi_H} \mathrm d \chi \frac 1{\chi^4}\, W_\gamma^3(\chi)\, D_+^4(a)
B_\delta^{\,\bk_1,\bk_2,\bk_3} \nn \\
T_\gamma^{\,\bl_1, \bl_2, \bl_3,\bl_4} &=& \int_0^{\chi_H} \mathrm d \chi \frac 1{\chi^6}\, W_\gamma^4(\chi)\, D_+^6(a)
T_\delta^{\,\bk_1,\bk_2,\bk_3,\bk_4}.
\ea 
Then, equivalent formulae as in eqn.~(\ref{eqn_source_spectra}) apply to these angular polyspectra, which are then related to the projected density field 
$\gamma(\bl)$ with two-dimensional angular wave vectors $\bl_i$:
\ba
\bra \gamma (\bl_1)\gamma (\bl_2)\gamma (\bl_3)\ket &=& (2 \pi)^3 \dirac(\bl_{1...3}) \, B_\gamma^{\,\bl_1, \bl_2, \bl_3} \nn \\
\bra \gamma (\bl_1)\gamma (\bl_2)\gamma (\bl_3)\gamma (\bl_4)\ket &=& (2 \pi)^3 \dirac(\bl_{1...4}) \, T_\gamma^{\,\bl_1, \bl_2, \bl_3,\bl_4} \, ,
\ea
where the fields on the sphere with angular directions $\bn_i$ are simply decomposed into Fourier harmonics instead of spherical harmonics
\ba
\bra \gamma(\bn_1)...\gamma(\bn_n) \ket &=& \int\frac{\mathrm{d}^2\ell_1}{(2\pi)^2}...\int\frac{\mathrm{d}^2\ell_n}{(2\pi)^2}\nn \\
		  ~&~&\cdot\, \bra \gamma(\bl_1)...\gamma(\bl_n) \ket \, \e ^{i \bl_1\cdot\bn_1}\,...\, \e^{i \bl_n\cdot\bn_n}\,.
\ea
Since the region on a sphere around a certain point can for small angles be approximated by the tangential plane, this is a good approximation 
for high $\ell$-values. It can generally be transformed to the full sky representation with Wigner 3j-symbols \citep{Hu2001}.
\subsection{Mixed iSW-galaxy polyspectra}\label{sect_notation}
An equivalent procedure of definitions as in the previous subsection can be applied to the iSW-fields $\tau(\bl)$. However, due to the uncorrelated 
noise sources in the iSW and galaxy fields mixed spectra are of predominant interest to us. If there exists a chance to securely measure the iSW 
signal it will only work via its cross-correlation to the projected galaxy density field $\gamma(\bl)$ in the cross power spectrum and higher order correlators. 

To allow a compact definition of the mixed spectra we introduce a doublet field $\ph_i(\bl)$
\be
\left(\begin{tabular}{r}$\ph_0(\bl)$\\
		    $\ph_1(\bl)$
		      \end{tabular}\right)= \left(\begin{tabular}{c}$\gamma\,(\bl)$\\
		    $\tau\,(\bl)$
		      \end{tabular}\right) \, .
\ee
Mixed spectra can now be defined in a compact way
\ba
\bra \ph_{i_1} (\bl_1)\ph_{i_2} (\bl_2)\ph_{i_3} (\bl_3)\ket &=& (2 \pi)^3 \dirac(\bl_{1...3}) \, B_{i_1i_2i_3}^{\,\bl_1, \bl_2, \bl_3}\nn \\
\bra \ph_{i_1} (\bl_1)\ph_{i_2}(\bl_2)\ph_{i_3} (\bl_3)\ph_{i_4} (\bl_4)\ket &=& (2 \pi)^3 \dirac(\bl_{1...4}) \, T_{i_1i_2i_3i_4}^{\,\bl_1,\bl_2,\bl_3,\bl_4}.
\ea
\subsection{Weighting functions}\label{sect_weighting}
\begin{figure}
\includegraphics[width=\columnwidth]{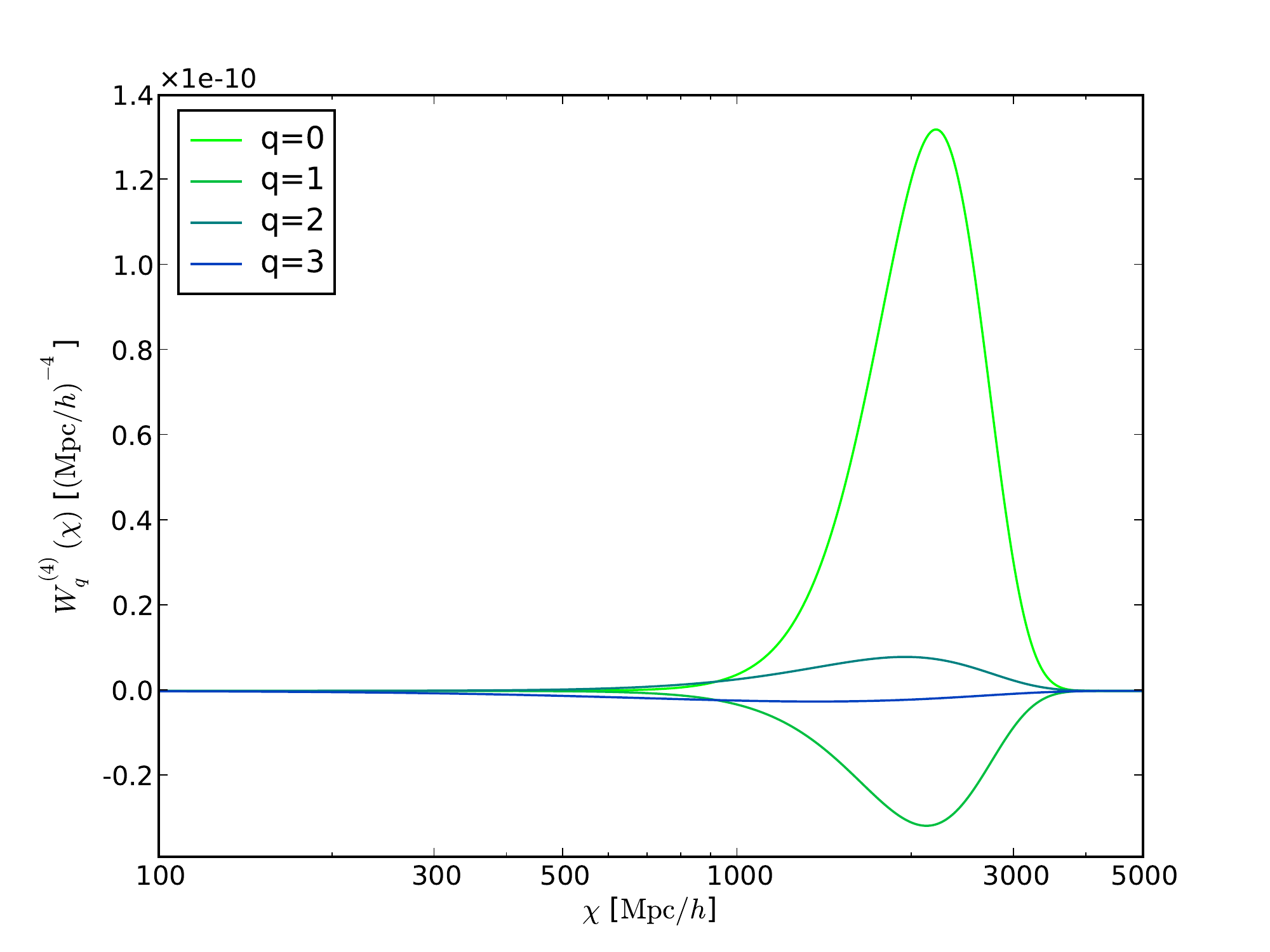}
\caption{Line-of-sight weighting functions $W_q^{(4)}(\chi)$ for mixed iSW-galaxy trispectra as a function of comoving distance.}
\label{fig_weighing_tri}
\end{figure}

For a mixed $n$-point function, the product of the $n$ different weighting functions, is uniquely given by the sum of the 
field indices $q$. In case of the bispectrum we would define $q=i_1+i_2+i_3$, whereas in case of the trispectrum $q=i_1+i_2+i_3+i_4$. 
We can therefore define a $q$-dependent combined weighting function $W_q^{(n)}(\chi)$
\be
W_q^{(n)}(\chi) = W_\tau^q(\chi) W_\gamma^{n-q}(\chi)\,
\ee
where $n=3$ and $n=4$ correspond to the bispectra and trispectra, respectively. The different weightings in case of the trispectra are 
depicted in Fig.~\ref{fig_weighing_tri} for different field mixtures $q$. Despite the weightings show strong differences in amplitude and sign, 
common to all weightings is a broad peak between 1 and 4 Gpc $h^{-1}$ due to the maximum in the galaxy redshift distribution $p(z)$.
\subsection{Time evolution}\label{sect_time_evo}
The time evolution of each linear galaxy field $\gamma(a,\bk)$ is given by the growth function $D_+(a)$. The $n$-th non-linear 
perturbative contributions evolve simply with the $n$-th order of the growth function $D_+^n$. This is not the case for the iSW field 
contributions $\tau(a,\bk)$. While the linear term evolves proportional to $\mathrm d (D_+/a)/\mathrm d a$ the higher orders can not just be written 
as the $n$-th power of the linear growth but are proportional to $\mathrm d (D_+^n/a)/\mathrm d a$.

Due to this fact, different perturbative contributions to mixed bispectra and trispectra will in general not have the same time evolution. 
In order to obtain a compact notation we introduce the time evolution doublet to $n$-th order $\bQ^{(n)}(a)$
\be
\bQ^{(n)}(a) = \left(\begin{tabular}{r}$Q_0^{(n)}(a)$\\
		    $Q_1^{(n)}(a)$
		      \end{tabular}\right)= \left(\begin{tabular}{c}$D_+^n$\\
		    $\frac{\mathrm d}{\mathrm d a}\left(\frac{D_+^n}{a}\right)$
		      \end{tabular}\right) \, .
\ee
With these time evolution functions $\bQ^{(n)}(a)$ we are now able to write down the general mixed time evolving source fields. For the tree-level bispectra we define 
\ba
B_{i_1 i_2 i_3}^{\,\bk_1,\bk_2,\bk_3} &=& (\chi_H\, k_1)^{-2 i_1}(\chi_H\, k_2)^{-2 i_2}(\chi_H\, k_3)^{-2 i_3}\nn \\
~&~& \left( \,\,~Q_{i_1}^{(2)}(a)\,Q_{i_2}^{(1)}(a)\,Q_{i_3}^{(1)}(a)\, b_\delta^{\,\bk_2,\bk_3}\right. \nn \\
~&~& \,\,+ Q_{i_2}^{(2)}(a)\,Q_{i_3}^{(1)}(a)\,Q_{i_1}^{(1)}(a)\, b_\delta^{\,\bk_3,\bk_1}\nn \\
~&~& \,\,\left.+ Q_{i_3}^{(2)}(a)\,Q_{i_1}^{(1)}(a)\,Q_{i_2}^{(1)}(a)\, b_\delta^{\,\bk_1,\bk_2}\,\,\right) \, .
\ea 
The terms $(\chi_H\, k_1)^{-2i}$ are the Poisson factors from the iSW effect. 
In case of the tree-level trispectrum the source will consist of two contributions - one originating from second order and third order perturbation theory 
respectively. The time dependent source for the trispectra then reads
\ba
T_{i_1 i_2 i_3 i_3}^{\,\bk_1,\bk_2,\bk_3 ,\bk_3}&=& (\chi_H\, k_1)^{-2 i_1}(\chi_H\, k_2)^{-2 i_2}(\chi_H\, k_3)^{-2 i_3}(\chi_H\, k_4)^{-2 i_4}\nn \\
~&~& \left(~ Q_{i_1}^{(2)}(a)\,Q_{i_2}^{(2)}(a)\,Q_{i_3}^{(1)}(a)\,Q_{i_4}^{(1)}(a)\, t_\delta^{(2)\,(\bk_1,\bk_2), (\bk_3,\bk_4)}\right. \nn \\
~&~& + \,\hbox{all pairs}\,\in\,\{1,2,3,4\}\nn \\
~&~& +\,  Q_{i_1}^{(3)}(a)\,Q_{i_2}^{(1)}(a)\,Q_{i_3}^{(1)}(a)\,Q_{i_4}^{(1)}(a)\, t_\delta^{(3)\,(\bk_1,\bk_2,\bk_3,\bk_4)}\nn \\
~&~& \left. + \,\mathrm{cyclic}\,\{1,\,2,\,3,\,4\}~~\right) \, .
\ea 
Now, the flat sky Limber equations for the mixed angular bispectra and trispectra 
read \citep{Hu2001}
\ba
B_{i_1i_2i_3}^{\,\bl_1, \bl_2, \bl_3} &=& \int_0^{\chi_H} \mathrm d \chi \frac 1{\chi^4}\, W_q^{(3)}(\chi)\,
B_{i_1i_2i_3}^{\,\bk_1,\bk_2,\bk_3} \nn \\
T_{i_1i_2i_3i_4}^{\,\bl_1, \bl_2, \bl_3,\bl_4} &=& \int_0^{\chi_H} \mathrm d \chi \frac 1{\chi^6}\, W_q^{(4)}(\chi)\, 
T_{i_1i_2i_3i_4}^{\,\bk_1,\bk_2,\bk_3,\bk_4},
\ea
where the source field spectra are evaluated at the 3-dimensional wave vectors $\bk_i  =(l_{i,1},l_{i,2},0) $. Since the weighting functions are slowly 
varying in comparison to the source field, fluctuations in the line-of-sight direction are smeared out by the integrations. Therefore, the fields can be assumed 
as non-fluctuating in this direction in the first place.

While pure spectra are invariant under exchange of wave vectors $\bl_i$,
\ba
B_{aaa}^{\,\bl_1, \bl_2, \bl_3}&=&B_{aaa}^{\,\bl_2,\bl_3,\bl_1}=B_{aaa}^{\,\bl_3,\bl_1,\bl_2} \nn \\
T_{aaaa}^{\,\bl_1, \bl_2, \bl_3, \bl_4}&=&T_{aaaa}^{\,\bl_2,\bl_3,\bl_4,\bl_1}=T_{aaaa}^{\,\bl_3,\bl_4,\bl_1,\bl_2}=T_{aaaa}^{\,\bl_4,\bl_1,\bl_2,\bl_3}\,,
\ea 
this does not generally hold true for mixed spectra $B_{abc}^{\,\bl_a,\bl_b,\bl_c}$. However, in general, all spectra are invariant under a 
simultaneous exchange of wave numbers and field indices
\ba
B_{i_1i_2i_3}^{\,\bl_1,\bl_2,\bl_3}&=&B_{i_2i_3i_1}^{\,\bl_2,\bl_3,\bl_1}=B_{i_3i_1i_2}^{\,\bl_3,\bl_1,\bl_2} \nn \\
T_{i_1i_2i_3i_4}^{\,\bl_1,\bl_2,\bl_3,\bl_4}&=&T_{i_2i_3i_4i_1}^{\,\bl_2,\bl_3,\bl_4,\bl_1}=T_{i_3i_4i_1i_2}^{\,\bl_3,\bl_4,\bl_1,\bl_2} =T_{i_4i_1i_2i_3}^{\,\bl_4,\bl_1,\bl_2,\bl_3}\, .
\ea
These symmetries are simply caused by the commutation invariance in the products of source fields in eqn.~(\ref{eqn_source_spectra}).
\subsection{Equilateral bispectra and square trispectra}\label{sect_equi}
\begin{figure}
\includegraphics[width=\columnwidth]{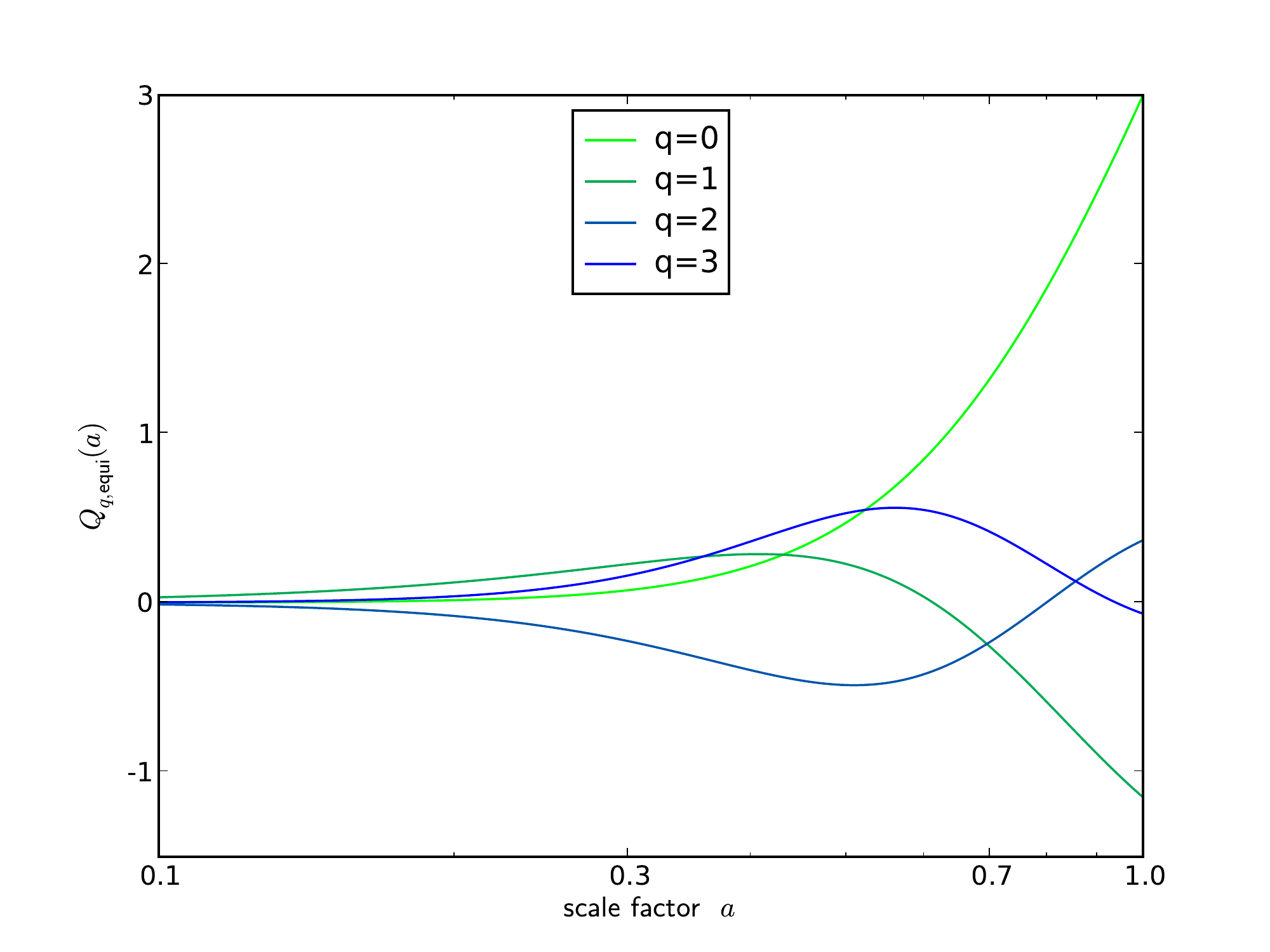}
\caption{Time evolution functions for equilateral mixed iSW-galaxy bispectra $B_{i_1i_2i_3}^{\,\ell,\ell,\ell}$ as a function of angular scale $\ell$. 
The value $q=i_1+i_2+i_3$ defines the mixture of the source fields. While the growth functions of the galaxy distribution stay positive to all perturbative orders, 
the derivatives in the iSW evolution functions also introduce negative terms into the evolution.}
\label{fig_Qbi_sym}
\end{figure}

To require homogeneity the wave vector arguments have to form a triangle, $\bl_1+\bl_2+\bl_3 = 0$, for the bispectrum and a quadrangle, 
$\bl_1+\bl_2+\bl_3+\bl_4 = 0$, in case of the trispectrum.
Thus, due to isotropy the scale dependence of the bispectrum is uniquely defined by the absolute values of the angular wave vectors $\bl_i$
\be
B_{abc}^{\,\bl_a,\bl_b,\bl_c} = B_{abc}^{\,\ell_a,\ell_b,\ell_c}\, .
\ee
The scale dependence of the trispectra, however, can be described by the four absolute values of the angular wave vectors $\bl_i$ and one diagonal $L$
\be
T_{abc}^{\,\bl_a,\bl_b,\bl_c, \bl_d} = T_{abc}^{\,\ell_a,\ell_b,\ell_c,\ell_d, L}\, .
\ee
This leads to the fact that the source fields of equilateral bispectra are symmetric with respect to their field indices and have a uniform time evolution 
$Q_{q,\mathrm{equi}}(a)$
\be
 Q_{q,\mathrm{equi}}=\left\{
\begin{tabular}{lcl}
 $D_+^3$ &$(q=0)$&\\
 $\frac13 D_+^2\left(\frac {\mathrm d}{\mathrm d a}\frac{D_+^2}{a}\right)+\frac23 D_+^3\left(\frac {\mathrm d}{\mathrm d a}\frac{D_+}{a}\right)$&$(q=1)$&\\
 $\frac23 D_+\left(\frac {\mathrm d}{\mathrm d a}\frac{D_+^2}{a}\right)  \left (\frac {\mathrm d}{\mathrm d a}\frac{D_+}{a}\right)
+ \frac13 D_+^2\left(\frac {\mathrm d}{\mathrm d a}\frac{D_+}{a}\right)^2$ &$(q=2)$ &\\
 $\left(\frac {\mathrm d}{\mathrm d a}\frac{D_+^2}{a}\right)\left(\frac {\mathrm d}{\mathrm d a}\frac{D_+}{a}\right)^2$ &$(q=3)$ &\,.\\
\end{tabular}\right .
\ee
These time evolutions are depicted in Fig.~\ref{fig_Qbi_sym}. While the growth functions of the galaxy distribution stay positive to all perturbative orders, 
the derivatives in the iSW evolution functions also introduce negative terms into the evolution. This will later lead to a change from correlation to 
anti-correlation along the line-of-sight. 

Slightly more complex is the time evolution for  source fields of the square trispectra. Here, the 
contributions from second order perturbation theory evolve still differently compared to the third order terms.
The second order terms $Q_{q,\mathrm{square}}^{(2)}(a)$ read
\begin{figure}
\includegraphics[width=\columnwidth]{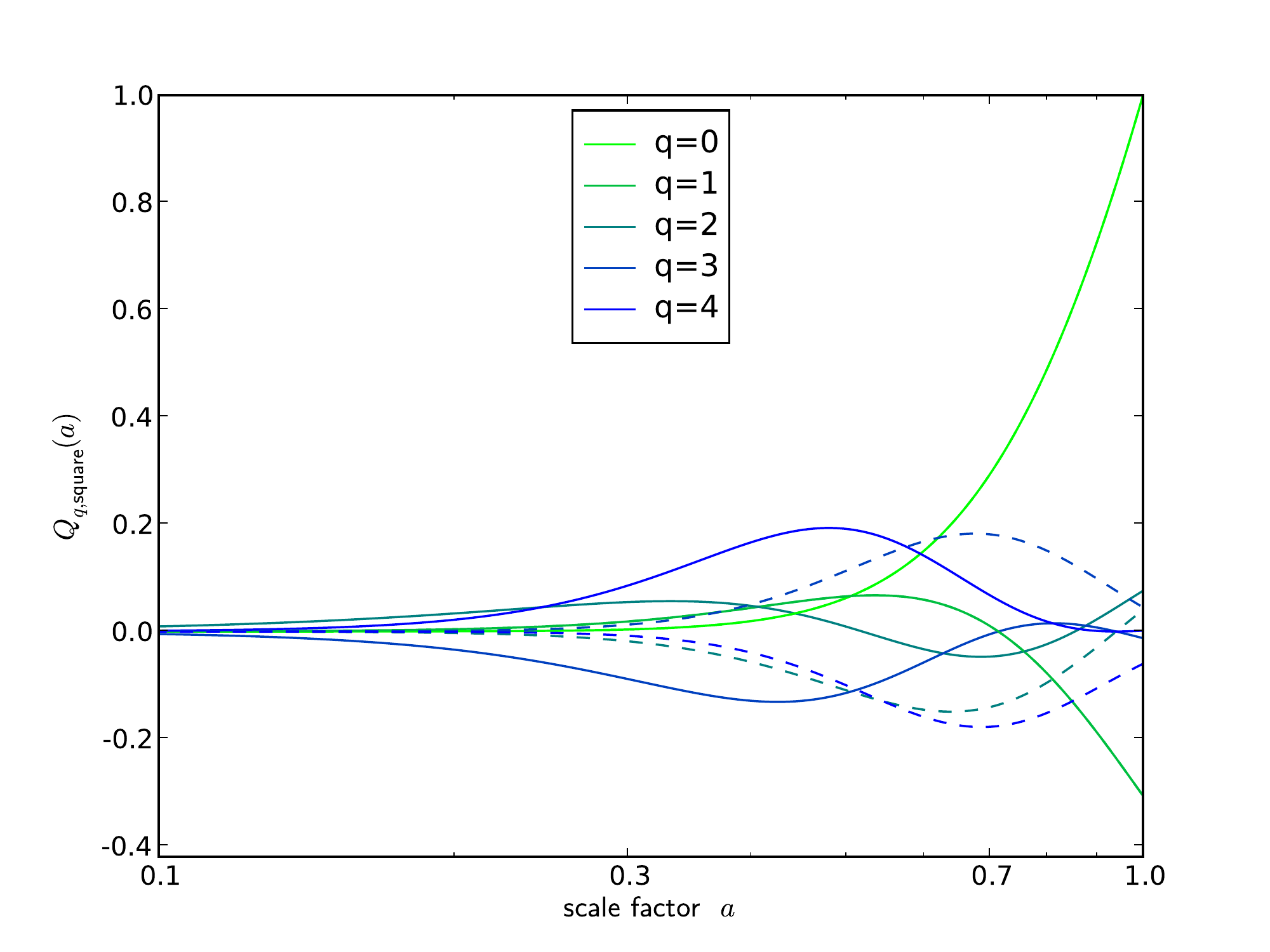}
\caption{Time evolution functions for square mixed iSW-galaxy trispectra $T_{(i_1i_2i_3i_4)}^{\, \ell,\ell,\ell,\ell,\sqrt2 \ell}$ as a function 
of angular scale $\ell$. 
The value $q=i_1+i_2+i_3+i_4$ defines the mixture of the source fields. The solid lines depict the second order perturbative time evolutions 
$Q_{q,\mathrm{square}}^{(2)}$, the third order terms $Q_{q,\mathrm{square}}^{(3)}$ are show as dashed lines. While the growth functions of the galaxy distribution stay positive to all perturbative orders, 
the derivatives in the iSW evolution functions also introduce negative terms into the evolution.}
\label{fig_Qtri_sym}
\end{figure}
\ba
Q_{q,\mathrm{square}}^{(2)}&=\left\{
\begin{tabular}{lcl}
 $D_+^6$ &$(q=0)$&\\
 $\frac12 D_+^4\left(\frac {\mathrm d}{\mathrm d a}\frac{D_+^2}{a}\right)+\frac12 D_+^5\left(\frac {\mathrm d}{\mathrm d a}\frac{D_+}{a}\right)$&$(q=1)$&\\
 $\frac 16 D_+^4\left(\frac {\mathrm d}{\mathrm d a}\frac{D_+}{a}\right)^2 + \frac23 D_+^3 \left(\frac {\mathrm d}{\mathrm d a}\frac{D_+^2}{a}\right)
\left (\frac {\mathrm d}{\mathrm d a}\frac{D_+}{a}\right)$ &~ &\\
  $ + \frac16 D_+^2\left(\frac {\mathrm d}{\mathrm d a}\frac{D_+^2}{a}\right)^2$ &$(q=2)$ &\\
 $\frac12 D_+\left(\frac {\mathrm d}{\mathrm d a}\frac{D_+^2}{a}\right)^2\left(\frac {\mathrm d}{\mathrm d a}\frac{D_+}{a}\right) 
+\frac12 D_+^2\left(\frac {\mathrm d}{\mathrm d a}\frac{D_+^2}{a}\right)\left(\frac {\mathrm d}{\mathrm d a}\frac{D_+}{a}\right)^2$&$(q=3)$&\\
 $\left(\frac {\mathrm d}{\mathrm d a}\frac{D_+^2}{a}\right)^2\left(\frac {\mathrm d}{\mathrm d a}\frac{D_+}{a}\right)^2$ &$(q=4)$ &
\end{tabular}\right. \nn \\
~&~\, \nn
\ea
and for the third order we obtain the following time evolution:
\ba
Q_{q,\mathrm{square}}^{(3)}&=\left\{
\begin{tabular}{lcl}
 $D_+^6$ &$(q=0)$&\\
 $\frac14 D_+^5\left(\frac {\mathrm d}{\mathrm d a}\frac{D_+}{a}\right)+\frac34 D_+^3\left(\frac {\mathrm d}{\mathrm d a}\frac{D_+^3}{a}\right)$&$(q=1)$&\\
 $\frac 12 D_+^4\left(\frac {\mathrm d}{\mathrm d a}\frac{D_+}{a}\right)^2 + \frac12 D_+^2\left(\frac {\mathrm d}{\mathrm d a}\frac{D_+^3}{a}\right)
\left(\frac {\mathrm d}{\mathrm d a}\frac{D_+}{a}\right)$ &$(q=2)$ &\\
 $\frac14 D_+^3\left(\frac {\mathrm d}{\mathrm d a}\frac{D_+^2}{a}\right)^3 
+\frac34 D_+\left(\frac {\mathrm d}{\mathrm d a}\frac{D_+^3}{a}\right)\left(\frac {\mathrm d}{\mathrm d a}\frac{D_+}{a}\right)^2$&$(q=3)$&\\
 $\left(\frac {\mathrm d}{\mathrm d a}\frac{D_+^3}{a}\right)\left(\frac {\mathrm d}{\mathrm d a}\frac{D_+}{a}\right)^3$ &$(q=4)$ &\,.
\end{tabular}\right.\nn \\
~&~\, \nn
\ea
In time evolutions for the different perturbative orders of the mixed trispectra are depicted in Fig.~\ref{fig_Qtri_sym}. One can see that for $q=1,2$ they 
evolve identically but differ stronger with increasing number $q$ of included iSW fields.

\begin{figure}
\includegraphics[width=\columnwidth]{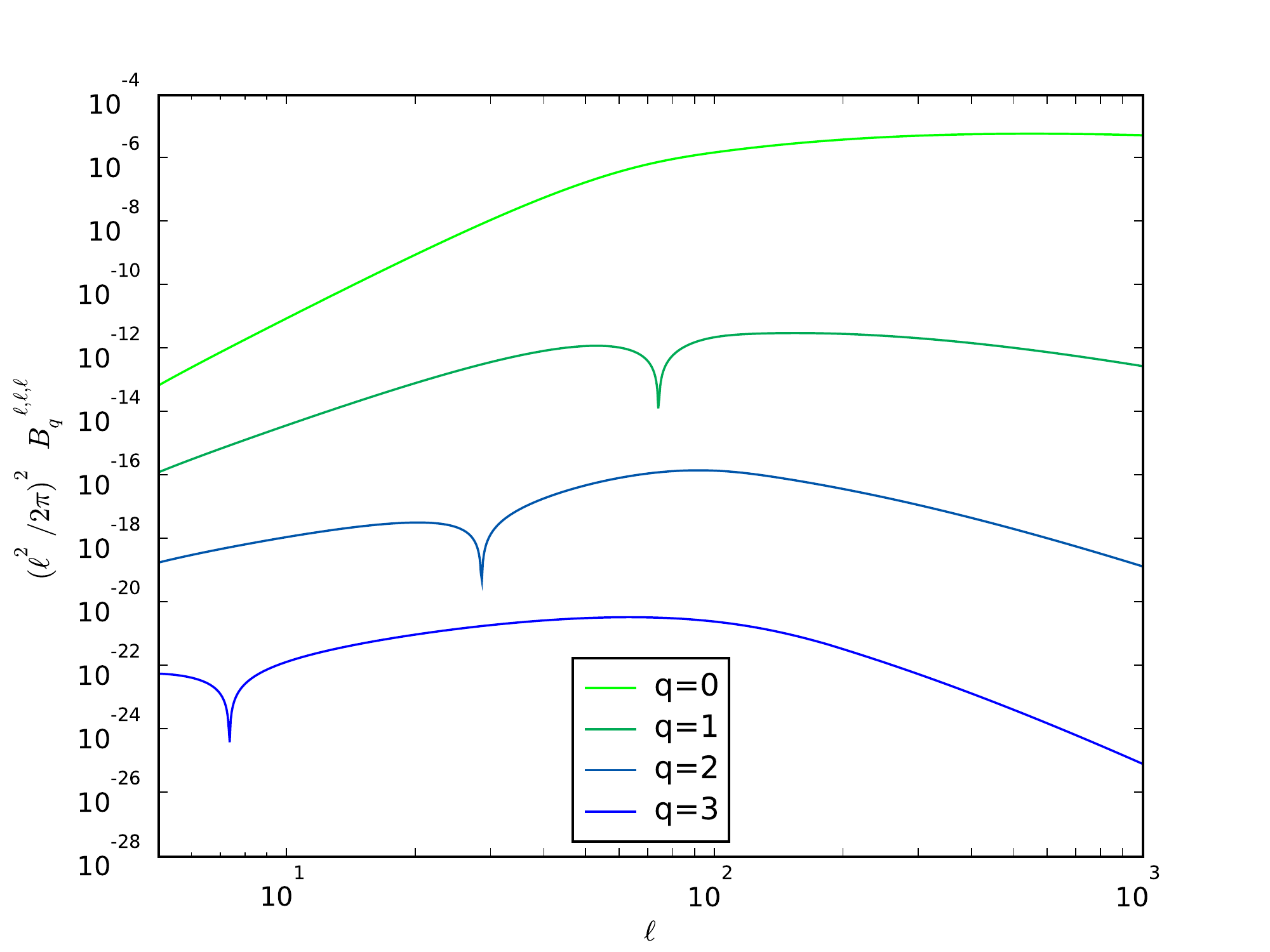}
\caption{The absolute values of the mixed equilateral iSW-galaxy bispectra $B_{i_1i_2i_3}^{\, \ell,\ell,\ell}$ as a function 
of angular scale $\ell$ are depicted in this figure. 
The value $q=i_1+i_2+i_3$ defines the mixture of the source fields.}
\label{fig_bi_equi}
\end{figure}

\begin{figure}
\includegraphics[width=\columnwidth]{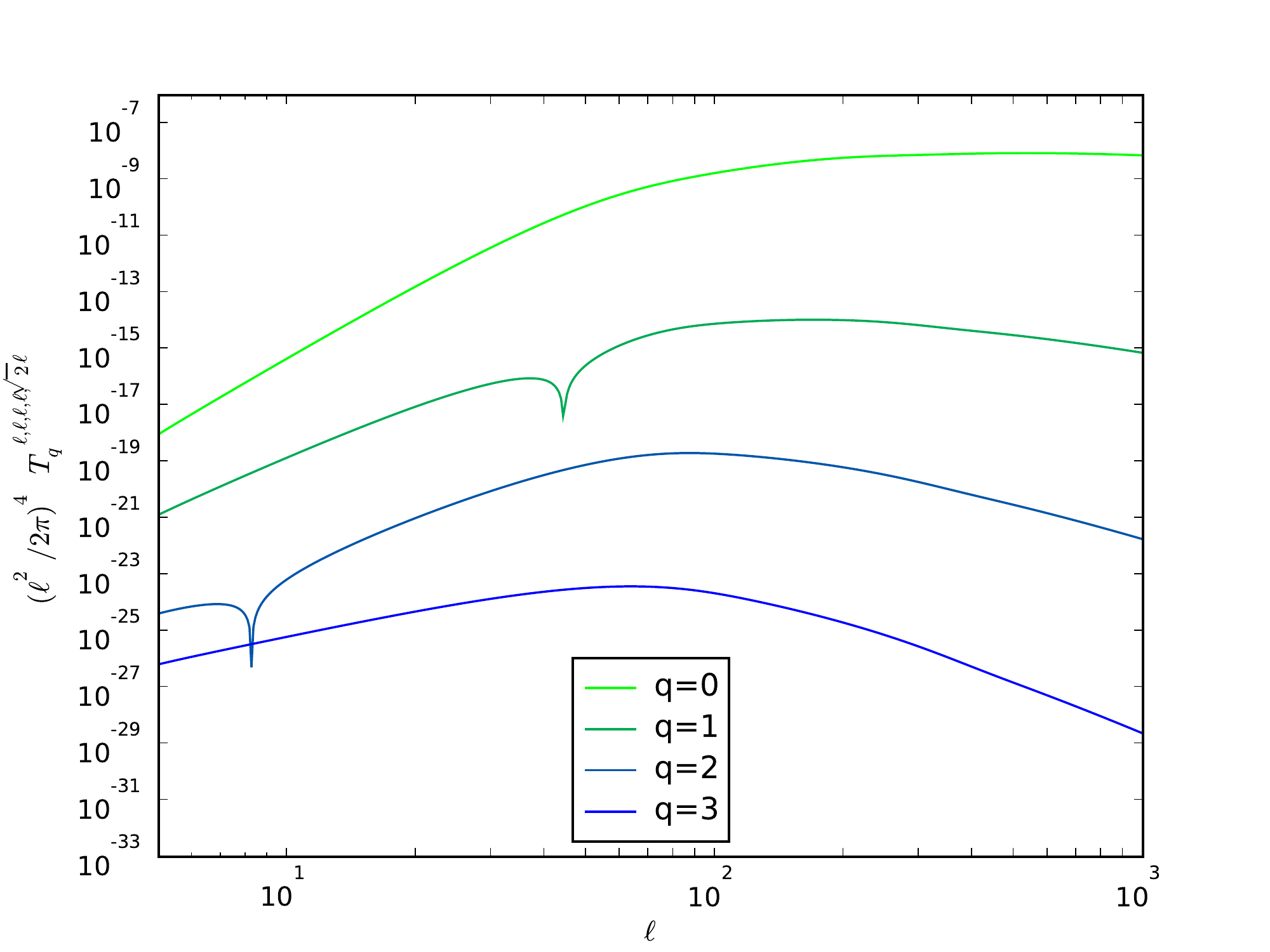}
\caption{The absolute values of the mixed square iSW-galaxy trispectra $T_{i_1i_2i_3i_4}^{\, \ell,\ell,\ell,\ell,\sqrt2 \ell}$ as a function 
of angular scale $\ell$ are shown in this plot. 
The value $q=i_1+i_2+i_3+i_4$ defines the mixture of the source fields.}
\label{fig_tri_quad}
\end{figure}

The equilateral bispectra and square trispectra are depicted in Fig.~\ref{fig_bi_equi} and Fig.~\ref{fig_tri_quad}. As in the power spectra one can observe also 
here the weakness of the iSW signal in comparison to the projected galaxy distribution field. This is clearly shown in the decrease of the polyspectra 
with increasing number $q$ of included iSW source fields. Once more, the iSW effect shows its nature of being a large scale effect. With higher $q$ the slope of 
the spectra increases in the large $\ell$ region. The physical reason for this is the coupling of the iSW effect to the gravitational potential in contrast to the 
galaxy distribution, which couples directly to the density contrast. Mathematically, this fact manifests itself in the appearance of the $1/k^2$ factors for the iSW 
contributions, originating from the inversion of the Poisson equation.  

While the linear iSW effect is strictly anti-correlated with respect to the galaxy density, cross-spectra with this linear signal would never show a change in sign 
in dependence on $\ell$. This does not hold true any longer for the non-linear iSW effect. The higher order contributions now have the opposite sign in their time 
evolution. It is therefore possible that for large $\ell$ the linear effect dominates while for small $\ell$ the non-linear effect determines the sign of the correlation. 
These changes in sign in dependence of $\ell$ can now be observed in Fig.~\ref{fig_bi_equi} and Fig.~\ref{fig_tri_quad}, for instance at $\ell \approx 80$ 
in the bispectrum $\bra \tau \gamma^2\ket$.
This behavior has also been observed and studied in the CMB-weak-lensing cross spectrum \citep{Nishizawa2008}.

\section{DETECTABILITY}\label{sect_detect}

\subsection{Sources of noise}\label{sect_noise}
The step from a good theoretical framework to an analysis of real data or to an estimation of the realistically accessible information content 
encompasses the description of all relevant effects influencing the measured data. Only then, one will be able to make statements about a physical process and 
the likelihood of its actual measurement. For the evaluation of the covariances of the bispectra and trispectra, the two-point function will be needed. In the same 
notation as the higher order spectra their are defined as
\be
 \bra \ph_{i_1}(\bl_1)\ph_{i_2}(\bl_2)\ket= (2\pi)^2 \dirac(\bl_1+\bl_2)\,C_{i_1i_2}^{\ell_1}
\ee
In the case at hand the actual theoretically expected iSW signal in our fiducial cosmological model is superposed to the primary CMB fluctuations. Its relative 
amplitude reaches from 10$\%$ on very large scales to an negligible fraction of the signal for scales smaller than $\ell\approx 200$. Furthermore, the detected CMB 
signal is subjected to instrumental noise $\sigma_\tau$ and a Gaussian beam $\beta(\ell)$.

Assuming that the noise sources of the galaxy counts are mutually uncorrelated, the pure galaxy-galaxy spectra are solely subjected to a Poissonian noise term $n^{-1}$.

The cross-spectra between the two fields will be free of noise, since the noise sources of the single fields are uncorrelated.   

Now we can relate the measured spectra $\tilde{C}_{i_1i_2}^{\ell}$ to the theoretical spectra $\tilde{C}_{i_1i_2}^{\ell}$:
\ba
\tilde{C}_{00}^{\ell}&=& C_{00}^{\ell} + n^{-1} \nn \\
\tilde{C}_{01}^{\ell}&=& C_{01}^{\ell}  \nn \\
\tilde{C}_{11}^{\ell}&=& C_{11}^{\ell} + C_{\mathrm{CMB}}^{\ell} + \sigma_\tau^2\,\beta^{-2}(\ell)\,.
\ea
The contributions in detail are:

(i) As the Fourier transform of the Gaussian beam one obtains $\beta^{-2}(\ell)= \exp(\Delta\theta^2\ell(\ell+1))$. We use $\Delta\theta = 7.1$ arcmin, 
which corresponds to the $\nu = 143 \,\,\mathrm{GHz}$ channels closest to the CMB emission maximum. For the conversion of $w_T^{-1} = T_\mathrm{CMB}^2\sigma_{\tau}^2$ 
to the noise amplitude in the dimensionless temperature perturbation $\tau$ with $w_T = (0.01\,\,\mu\mathrm K)^2$ \citep{Zaldarriaga1997} we use the 
value $T_\mathrm{CMB} = 2.725 \,\,\mathrm K $ for the CMB temperature.
\begin{table}
\label{tab_euclid}
\centering
\begin{tabular}{cccccc}
\hline
$N$		&	$\Delta \Omega$	& $f_\mathrm{sky} $ 	&	$z_0$  	&	$b$	&	$n$\\
\hline	
$3.0 \times10^9$&	$2\pi$		&$0.5$			&$0.64$		&$1.0$ 		&$4.7\times 10^8$\\
\hline
\end{tabular}
\caption{The Properties of the \textit{Euclid} galaxy survey are listed in this table: total number $N$ of objects, solid angle $\Delta\Omega$ 
covered (in radians), sky fraction $f_\mathrm{sky}$, redshift parameter $z_0$, galaxy bias $b$ and density per unit steradian $n$.}
\end{table}

\begin{figure}
\includegraphics[width=\columnwidth]{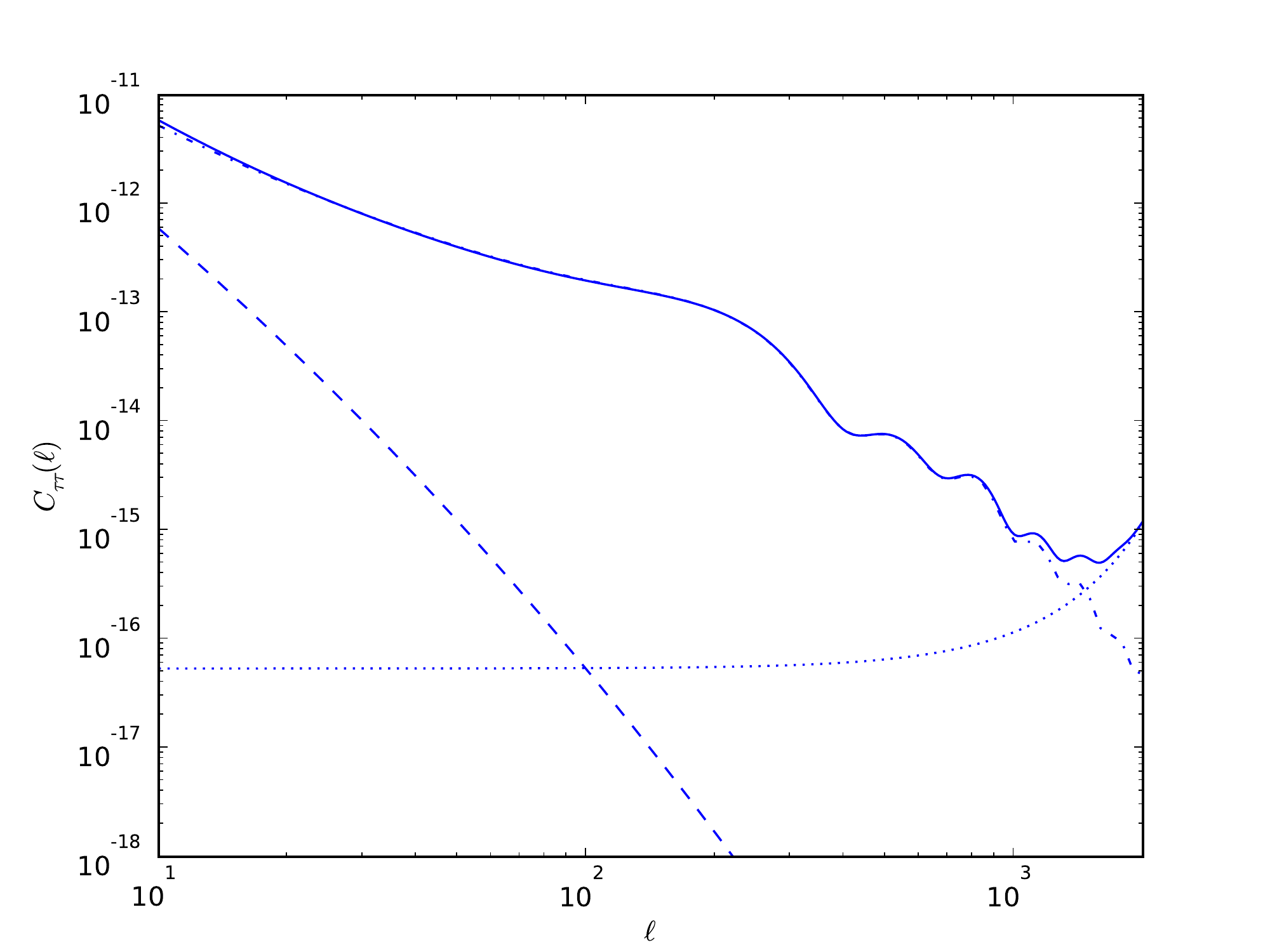}
\caption{Constituents of the measured angular CMB spectrum $\tilde{C}_{11}^{\ell}$. Depicted are the total signal $\tilde{C}_{11}^{\ell}$ (solid line), 
the contribution from primordial fluctuations $ C_{\mathrm{CMB}}^{\ell}$, the iSW-effect $C_{11}^{\ell}$ and the instrumental noise 
$\sigma_\tau^2\,\beta^{-2}(\ell)$, which is fortunately sub-dominant at the large scales of interest.}
\label{fig_c_tt}
\end{figure}

(ii) Furthermore, a CMB temperature power spectrum $C_\mathrm{CMB}(\ell)$ was generated, which was equally scaled with the CMB temperature 
$T_\mathrm{CMB} = 2.725 \,\,\mathrm K$, with the Code for Anisotropies in the Microwave Background \citep[CAMB,][]{Lewis2000} for the fiducial $\Lambda$CDM 
cosmology. The noise contribution from the CMB-spectrum $C_{\mathrm{CMB}}^{\ell}$ represents the main challenge in the observation of the iSW bispectra and trispectra. 
It provides high values for the covariance at low multipoles $\ell$, and it by far dominates $\tilde C _{00}(\ell)$, $C_\mathrm{CMB}\gg C_{00}(\ell)$ on the 
angular scales considered. The orders of magnitude for the different contributions of the linear estimator $\tilde{C}_{11}^{\ell}$ are depicted in Fig~\ref{fig_c_tt}. 
One can see, that even at large angular scales $\ell$ the pure iSW signal $C_{11}^{\ell}$ is still more than one order of magnitude weaker than the signal from 
primordial fluctuations $ C_{\mathrm{CMB}}^{\ell}$.

(iii) The inverse number density $n$ of objects per unit steradian determines the Poissonian noise term in the galaxy count. In Table~\ref{tab_euclid} 
the properties of the main galaxy sample as it would be expected from \textit{Euclid} are summarized. 
Major advantages lie in the large sky coverage and the high number of observed objects. Here, we assumed a non-evolving galaxy bias for simplicity.
\subsection{Covariances}\label{sect_cov}
In the case of Gaussian noise the observed and estimated bispectra $\tilde B_{i_1i_2i_3}^{\,\ell_1,\ell_2,\ell_3}$ and the trispectra 
$\tilde T_{i_1i_2i_3i_4}^{\,\bl_1,\bl_2,\bl_3,\bl_4}$ are unbiased estimates of the true bispectra $B_{i_1i_2i_3}^{\,\ell_1,\ell_2,\ell_3}$ 
and trispectra $T_{i_1i_2i_3i_4}^{\,\bl_1,\bl_2,\bl_3,\bl_4}$ \citep{Hu2001},
\ba
\tilde B_{i_1i_2i_3}^{\,\ell_1,\ell_2,\ell_3}&\simeq&B_{i_1i_2i_3}^{\,\ell_1,\ell_2,\ell_3}\nn \\
\tilde T_{i_1i_2i_3i_4}^{\,\bl_1,\bl_2,\bl_3,\bl_4}&\simeq&T_{i_1i_2i_3i_4}^{\,\bl_1,\bl_2,\bl_3,\bl_4}\,.
\ea
This is in contrast to the spectra $C_{i_1i_2}^{\ell}$, which were discussed in the previous subsection. The covariances of the estimators of the 
bispectra and trispectra are defined as
\ba
\mathrm{Cov}\left[\tilde B_{i_1i_2i_3}^{\,\ell_1,\ell_2,\ell_3}, \tilde B_{i_1i_2 i_3}^{\,\ell_1^\prime,\ell_2^\prime,\ell_3^\prime}\right]\hspace{3cm} &~ \nn \\
=\left \bra \left( \tilde B_{i_1i_2i_3}^{\,\ell_1^{~},\ell_2,\ell_3} - B_{i_1i_2i_3}^{\,\ell_1,\ell_2,\ell_3} \right) \right.
\left. \left( \tilde B_{i_1i_2i_3}^{\,\ell_1^\prime,\ell_2^\prime,\ell_3^\prime} 
- B_{i_1i_2i_3}^{\,\ell_1^\prime,\ell_2^\prime,\ell_3^\prime} \right)\right \ket\,, &  ~\nn  
\ea
\ba
\mathrm{Cov}\left[\tilde T_{i_1i_2i_3,i_4}^{\,\bl_1,\bl_2,\bl_3,\bl_4}, 
\tilde T_{i_1i_2 i_3, i_4}^{\,\bl_1^\prime,\bl_2^\prime,\bl_3^\prime, \bl_4^\prime}\right]\hspace{4cm}&~ \nn \\
=\left \bra \left( \tilde T_{i_1i_2i_3,i_4}^{\,\ell_1^{~},\ell_2,\ell_3,\ell_4, L} -  T_{i_1i_2i_3,i_4}^{\,\bl_1,\bl_2,\bl_3,\bl_4}\right) \right.
\left.\left( \tilde T_{i_1i_2 i_3, i_4}^{\,\bl_1^\prime,\bl_2^\prime,\bl_3^\prime, \bl_4^\prime}
-T_{i_1i_2 i_3, i_4}^{\,\bl_1^\prime,\bl_2^\prime,\bl_3^\prime, \bl_4^\prime}\right)\right \ket \,.& ~
\ea
In a Gaussian approximation, which we are using here, any covariances can be expressed as a sum of products of two-point functions using Wick's theorem. 
While for pure covariances 
only the respective power spectra appear in this expansion, in our case of mixed covariances the products are formed from the estimators of the cross-correlation 
$\tilde C_{01}^{\ell}$ and the two auto-correlations $\tilde C_{00}^{\ell}$ and $\tilde C_{11}^{\ell}$. 

In case of the bispectra with mutually unequal angular wave numbers $\ell_{i_j}\neq\ell_{i_k}$ for $j\neq k$ the covariance can be written as a sum over 
terms which are cubic in the spectra $C_{i_1i_2}^{\ell}$
\begin{multline}
\mathrm{Cov}\left[\tilde B_{abc}^{\,\ell_1,\ell_2,\ell_3}, \tilde B_{a^\prime b^\prime c^\prime}^{\,\ell_1^\prime,\ell_2^\prime,\ell_3^\prime}\right]\\
= \tilde C_{aa^\prime}^{\ell_1}\tilde C_{bb^\prime}^{\ell_2} \tilde C_{cc^\prime}^{\ell_3}\,
\,\dirac(\bl_1^{~}-\bl_1^\prime) \,\dirac(\bl_2^{~}-\bl_2^\prime)\,\dirac(\bl_3^{~}-\bl_3^\prime) \\
+\mathrm{perm}(\bl_1^\prime, \bl_2^\prime ,\bl_3^\prime)\,.
\end{multline}
On the subspace $\ell_1,\ell_1^\prime < \ell_2,\ell_2^\prime < \ell_3\ell_3^\prime$ only the first term is non-vanishing. This block-diagonal matrix can now 
be inverted to
\begin{multline}
\label{cov_inv_bi}
\mathrm{Cov}^{-1}\left[\tilde B_{abc}^{\,\ell_1,\ell_2,\ell_3}, \tilde B_{a^\prime b^\prime c^\prime}^{\,\ell_1^\prime,\ell_2^\prime,\ell_3^\prime}\right]
= \frac{\tilde C_{aa^\prime}^{\ast\,\ell_1}\tilde C_{bb^\prime}^{\ast\,\ell_2}\tilde C_{cc^\prime}^{\ast\,\ell_3} }
{\det C^{\,\ell_1}\det C^{\,\ell_2}\det C^{\,\ell_3}}  \\ \times \,\dirac(\bl_1^{~}-\bl_1^\prime) \,\dirac(\bl_2^{~}-\bl_2^\prime)\,\dirac(\bl_3^{~}-\bl_3^\prime)\,,
\end{multline}
with the adjoint matrix $\tilde C_{aa^\prime}^{\ast\,\ell}$,
\be
\tilde C^{\ast\,\ell} = \left(
\begin{tabular}{rr}
$C_{11}^{\,\ell}$&$-C_{01}^{\,\ell}$\\
$-C_{01}^{\,\ell}$&$C_{00}^{\,\ell}$
\end{tabular}\right)\,.
\ee
In analogy to the bispectrum case, the inverse covariance of the trispectra in the subspace 
$\ell_1,\ell_1^\prime < \ell_2,\ell_2^\prime < \ell_3,\ell_3^\prime <\ell_4,\ell_4^\prime$ amounts to
\begin{multline}
\mathrm{Cov}^{-1}\left[\tilde T_{abcd}^{\,\bl_1,\bl_2,\bl_3,\bl_4}, \tilde T_{a^\prime b^\prime c^\prime d^\prime}
^{\,\bl_1^\prime,\bl_2^\prime,\bl_3^\prime,\bl^\prime}\right] 
= 
\frac{\tilde C_{aa^\prime}^{\ast\,\ell_1}\tilde C_{bb^\prime}^{\ast\,\ell_2}\tilde C_{cc^\prime}^{\ast\,\ell_3}\tilde C_{dd^\prime}^{\ast\,\ell_4}}
{\det C^{\,\ell_1}\det C^{\,\ell_2}\det C^{\,\ell_3}\det C^{\,\ell_4}} \\
\times\,\dirac(\bl_1^{~}-\bl_1^\prime)\, \dirac(\bl_2^{~}-\bl_2^\prime)\,\dirac(\bl_3^{~}-\bl_3^\prime)\,\dirac(\bl_4^{~}-\bl_4^\prime)\,.
\end{multline}
For an observation covering the sky with a fraction of $f_\mathrm{sky}$ the covariances scale like $f_\mathrm{sky}^{-1}$. The anti-correlation in the cross-spectra 
$C_{01}^{\ell}$ will not change the sign of the covariances, since in each of the products an even number  of these mixed spectra appears.
\subsection{Signal-to-noise ratios}\label{sect_s2n}
The signal-to-noise ratio $\Sigma^{(3)}$ for the simultaneous measurements of the all pure and mixed bispectra $\bra\tau^q\gamma^{3-q}\ket$ and $\Sigma_q^{(4)}$ 
for the all mixed and pure trispectra $\bra\tau^q\gamma^{4-q}\ket$, where all field indices are summed over, 
would imply a thorough derivation of all cross-correlations between different field mixtures. Here, we are rather interested in the individual signal-to-noise ratios 
of certain field configurations. 
If one reduces the data to a mixed configuration $q$, only the cases $q=0$ and $q=1$ provide measurement in and above the detection limit. 
For $q=0$ we obtain
\ba
\left( \Sigma_0^{(3)} \right)^2 &=& \frac {f_\mathrm{sky}}{4\pi^3}
      \int \mathrm d^2 \bl_{1}\, \mathrm d^2 \bl_{2}\,\,
       \frac{\left(B_{000}^{\,\ell_1,\ell_2,\ell_3}\right)^2}{6\,\tilde C_{00}^{\,\ell_1}\tilde C_{00}^{\,\ell_2}\tilde C_{00}^{\,\ell_3}} \nn \\
\left( \Sigma_0^{(4)} \right)^2 &=& \frac {f_\mathrm{sky}}{8\pi^4} 
      \int \mathrm d^2 \bl_{1}\, \mathrm d^2 \bl_{2}\, \mathrm d^2 \bl_{3}\,\,
       \frac{\left(T_{0000}^{\,\ell_1,\ell_2,\ell_3,\ell_4}\right)^2}{24\,\tilde C_{00}^{\,\ell_1}\tilde C_{00}^{\,\ell_2}\tilde C_{00}^{\,\ell_3}\tilde C_{00}^{\,\ell_4}}
\ea
However, since we are aiming for iSW detections, the more interesting case is $q=1$. The signal-to-noise ratio then splits up into two contributions,
\ba
\label{s2n_bi_q1}
\left( \Sigma_1^{(3)} \right)^2 &=& \frac {f_\mathrm{sky}}{4\pi^3}
      \int \mathrm d^2 \bl_{1}\, \mathrm d^2 \bl_{2}\,\,\left(2\,\det C^{\,\ell_1}\det C^{\,\ell_2}\det C^{\,\ell_3}\right)^{-1}\nn\\
       ~&~& \left[\left(B_{001}^{\,\ell_1,\ell_2,\ell_3}\right)^2\tilde C_{11}^{\,\ell_1}\tilde C_{11}^{\,\ell_2}\tilde C_{00}^{\,\ell_3}+B_{001}^{\,\ell_1,\ell_2,\ell_3}\tilde C_{11}^{\,\ell_1}\tilde C_{01}^{\,\ell_2}\tilde C_{01}^{\,\ell_3} B_{010}^{\,\ell_1,\ell_2,\ell_3}\right]\nn \\
\left( \Sigma_1^{(4)} \right)^2 &=& \frac {f_\mathrm{sky}}{8\pi^4}
      \int \mathrm d^2 \bl_{1}\, \mathrm d^2 \bl_{2}\, \mathrm d^2 \bl_{3}\,\,\left(\det C^{\,\ell_1}\dots\det C^{\,\ell_4}\right)^{-1}\nn\\
       ~&~& \left[\,\,\,\frac{1}{6}\,\left(T_{0001}^{\,\ell_1,\ell_2,\ell_3,\ell_4}\right)^2 \tilde C_{11}^{\,\ell_1}\tilde C_{11}^{\,\ell_2}\tilde C_{11}^{\,\ell_3}\tilde C_{00}^{\,\ell_4}\right. \nn \\
       ~&~& +\,\frac{1}{4}\,\left.T_{0001}^{\,\ell_1,\ell_2,\ell_3,\ell_4}\tilde C_{11}^{\,\ell_1}\tilde C_{11}^{\,\ell_2}\tilde C_{01}^{\,\ell_3}\tilde C_{01}^{\,\ell_4}T_{0010}^{\,\ell_1,\ell_2,\ell_3,\ell_4}\,\right]\,.
\ea
A detailed calculation of the signal-to noise expressions can be found in Section~\ref{ap1}. 
The inverse covariances of the polyspectra will always remain positive, since always an even number of 
anti-correlating cross-spectra will appear in its expression. However, the mixed field contributions can in general become negative.
\begin{figure}
\includegraphics[width=\columnwidth]{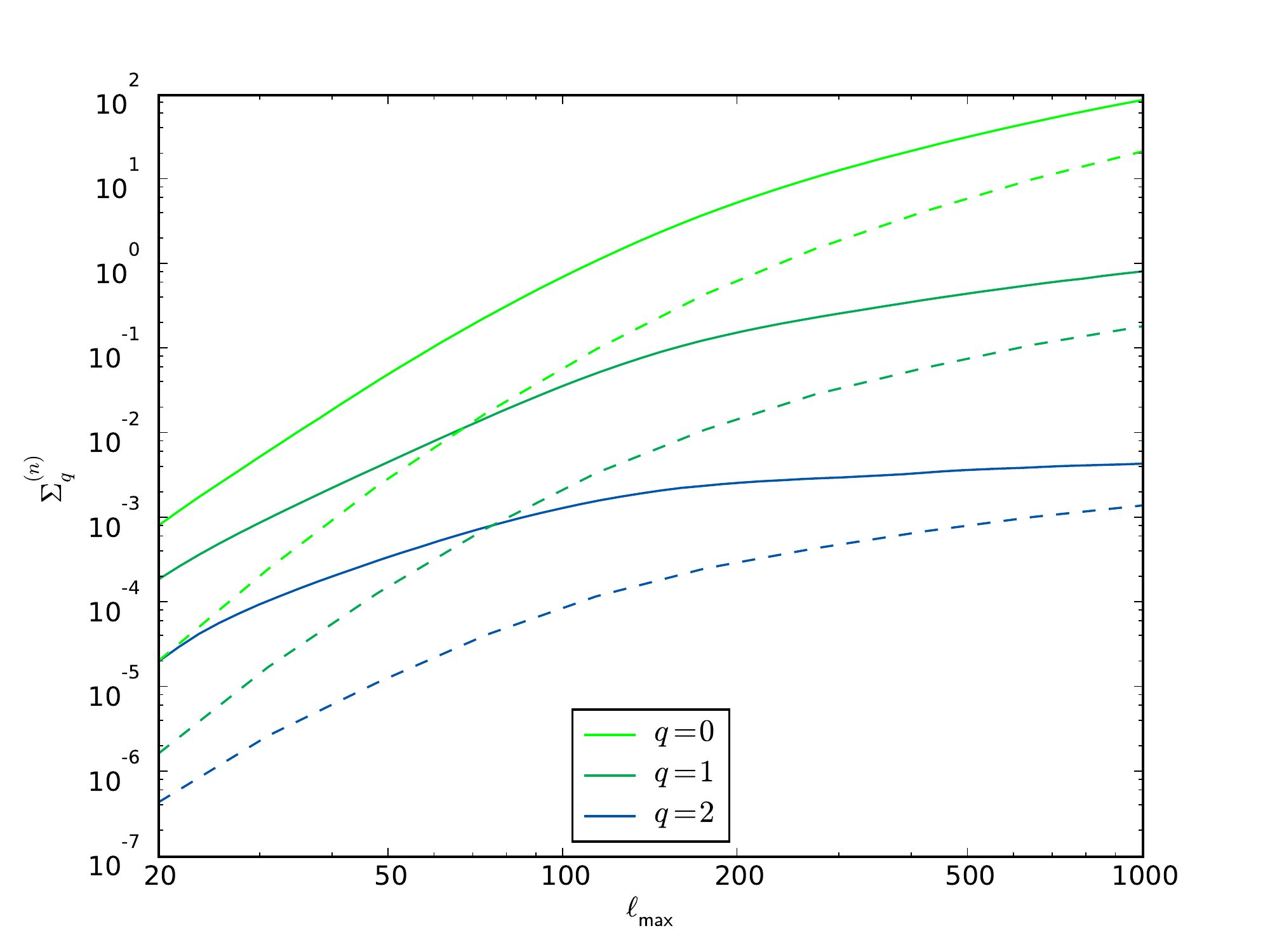}
\caption{Cumulative signal-to-noise ratios $\Sigma_q^{(n)}$ for measurements of the bispectra $\bra \tau^q \gamma^{3-q}\ket$ (solid lines) 
and the trispectra $\bra \tau^q \gamma^{4-q}\ket$ (dashed lines), 
$q= 0,1$, for \textit{Planck} CMB data in cross-correlation with \textit{Euclid}-like survey, up to a 
resolution limit $\ell_\mathrm{max} = 10^3$ starting from a minimum angular wave number of $\ell_\mathrm{min}=10$.}
\label{fig_s2n_tri_bi}
\end{figure}
\begin{table}
\label{tab_s2n}
\centering
\begin{tabular}{c|ccc}
\hline
$q$		&	$0$	& 	$1$ 	&	$2$  	\\
\hline	
$\Sigma_q^{(3)}$&	$87.8$	&	$0.828$	&$4.43\cdot 10^{-3}$	\\
$\Sigma_q^{(4)}$&	$21.7$	&	$0.19$	&$1.42\cdot 10^{-3}$	\\
\hline
\end{tabular}
\caption{Cumulative signal-to-noise ratios $\Sigma_q^{(n)}$ for measurements of the bispectra $\bra \tau^q \gamma^{3-q}\ket$ and the 
trispectra  $\bra \tau^q \gamma^{4-q}\ket$, $q= 0,1,2$, for \textit{Planck} CMB data in cross-correlation with \textit{Euclid}-like survey, up to a 
resolution limit $\ell_\mathrm{max} = 10^3$ starting from a minimum angular wave number of $\ell_\mathrm{min}=10$.}
\end{table}

The cumulative signal-to-noise ratios $\Sigma_q^{(n)}$ for the mixed bispectra $B_q$ and the mixed trispectra $T_q$ are depicted in Fig.~\ref{fig_s2n_tri_bi} for 
the pure galaxy spectra and spectra with up to two iSW source fields included, $q=0,1,2$. 

The qualitative behavior of the cumulative signal-to-noise curves are again determined by the individual signal strengths of the two source fields 
$\gamma$ and $\tau$. The strong fluctuation of the galaxy distribution $\gamma$ even on small scales leads to a considerable increase of $\Sigma$ for large $\ell$ 
and small $q$. In contrast to this the iSW-effect is a large scale effect and therefore increases the slope in the small $\ell$ limit of the spectrum. 
It does not contribute significant signal strength above values of $\ell_\mathrm{max}>300$, for this reason the signal-to-noise curves flatten off in this 
region of the spectrum for $q=1,2$. The wider spread between different values of $q$ for the trispectrum in contrast to the bispectrum is due to the higher power of 
source fields.

Quantitatively, higher values of $q$ lead to smaller significance in the signal. Included were contributions starting from large angular scales 
$\ell_\mathrm{min}=10$ up to smallest scales measurable in the \textit{Planck} survey $\ell_\mathrm{max}= 10^3$. The pure galaxy polyspectra 
$\bra \gamma^3\ket$ and $\bra \gamma^4\ket$ can both be measured with a detection significance of $\gg3\sigma$, $\Sigma_0^{(3)}=87.8$ and $\Sigma_0^{(4)}=21.7$. 
Including only one iSW source field reduces the signal down to the noise level. 
While the bispectrum $\bra \tau \gamma^2 \ket$ reaches a signal-to-noise ratio of $0.82$, the value for the trispectrum 
$\bra \tau \gamma^3 \ket$ reaches a maximum of $0.19$. Combining measurements of the $q=1$ bi- and trispectra would therefore be able to contribute 
a maximum signal-to-noise contribution of $\Sigma\approx 0.84$. Unfortunately, this is - taken on its own - still a very poor measurement significance. 
However, it could be used as an additional signal source to the strongest iSW signal from the cross spectrum $\bra \tau \gamma\ket$.

For the higher values of $q$ only the case of two iSW source fields $q=2$ is plotted in Fig.~\ref{fig_s2n_tri_bi}. Both for the bispectrum as well as for the 
trispectra the signal-to-noise ratios are negligible with maximum values of $4.43\cdot 10^{-3}$ and $1.42\cdot 10^{-3}$ respectively.

One can obtain a grasp of the differential contributions of the signal-to-noise ratios with respect to angular scale $\ell$, if one studies the quantity
\ba
\left(\frac{\mathrm d}{\mathrm d \ell}\,\Sigma_{q,\mathrm{equi}}^{(3)} \right)^{\frac12} &\propto& 
B_q^{\ell,\ell,\ell}\sqrt{\mathrm{Cov}^{-1}\left(B_q^{\ell,\ell,\ell}\right)} \nn \\
\left(\frac{\mathrm d}{\mathrm d \ell}\,\Sigma_{q,\mathrm{square}}^{(4)} \right)^{\frac12} &\propto& 
T_q^{\ell,\ell,\ell,\ell,\sqrt{2}\ell}\sqrt{\mathrm{Cov}^{-1}\left(T_q^{\ell,\ell,\ell,\ell,\sqrt{2}\ell}\right)} \,.
\ea
\begin{figure}
\includegraphics[width=\columnwidth]{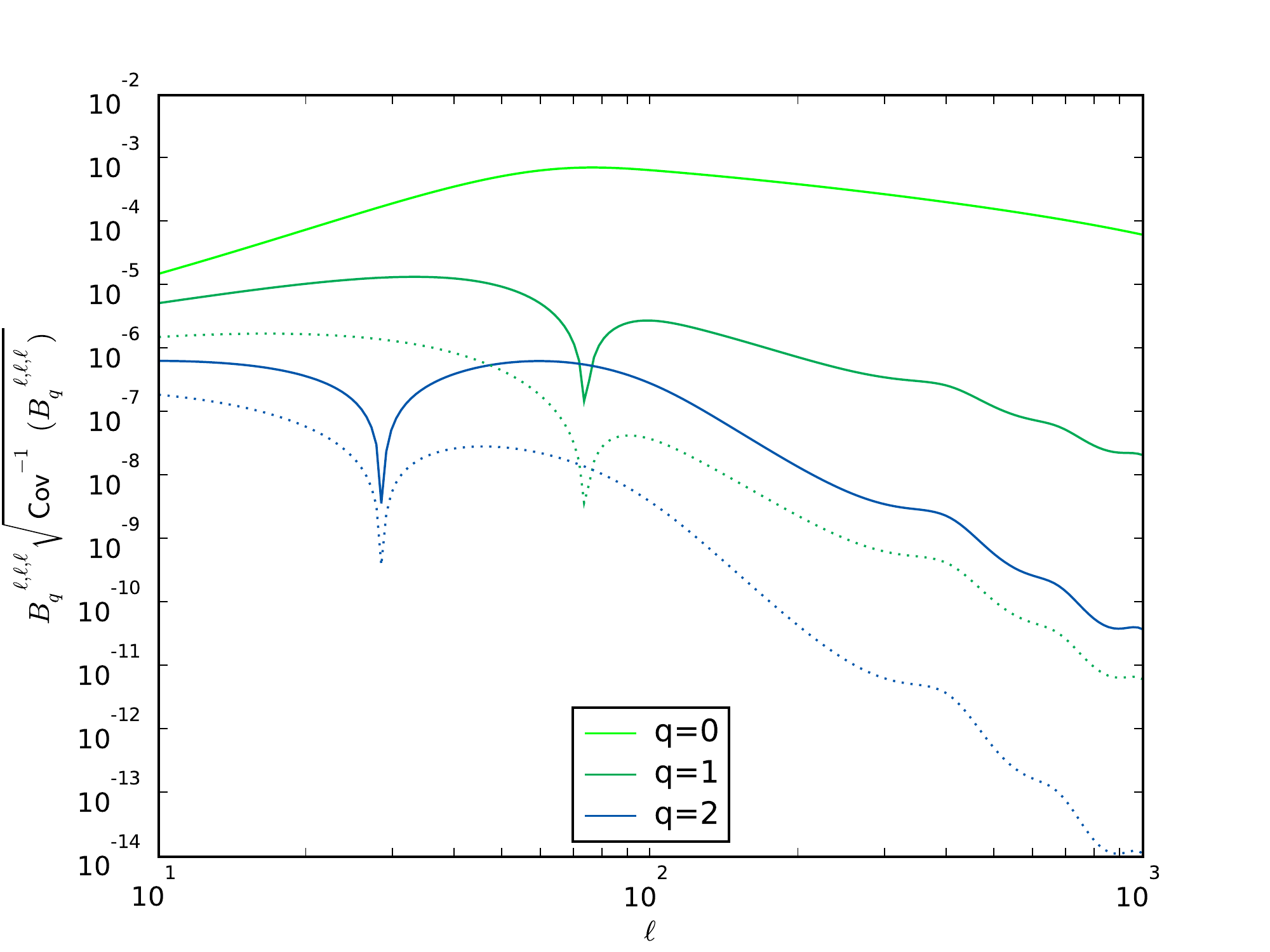}
\caption{The differential contributions of the equilateral bispectra to the signal-to-noise ratios in dependence on angular wave number $\ell$ are depicted 
here for different source field mixtures $q=0,1,2$ (solid lines). For $q=1,2$ the contributions from cross-correlations are also shown (dotted lines), as they appear 
for $q=1$ in the second term of the second line in eqn.~(\ref{s2n_bi_q1}). 
One can observe the increasing amplitude of the baryonic acoustic oscillations for larger $q$. 
Also the change in sign can be studied due to the transition from linear dominated to non-linear dominated scales.}
\label{fig_bi_equi_diff}
\end{figure}
In Fig.~\ref{fig_bi_equi_diff} this differential contribution of equilateral bispectra in dependence on $\ell$ are depicted for different source field 
mixtures $q$. The differential contributions of the square trispectra behave qualitatively analogous.
Also the change in sign can be studied due to the transition from linear dominated to non-linear dominated scales. For $q=1,2$ the contributions from cross-correlations 
are also shown (dotted lines), as they appear for $q=1$ in the second term of the second line in eqn.~(\ref{s2n_bi_q1}). As one can see, these terms are subdominant 
and can be neglected in our case.

One can observe the increasing amplitude of the baryonic acoustic oscillations for larger $q$, which originate from $C_\mathrm{CMB}^\ell$. 
The falling slopes of the BAO features in the covariance lead to 
small plateaus in the differential contributions for larger $\ell$. Since in these regions the signal decreases more gently than the covariance, one obtains a 
local increase of signal-to-noise. However, this effect can hardly be observed in Fig.~\ref{fig_s2n_tri_bi}.
\section{SUMMARY}\label{sect_sum}
The objective of this work is a study of the detectability of non-Gaussian signatures in non-linear iSW-effect. Besides the mixed bispectra of the form 
$\bra \tau^q \gamma^{3-q}\ket$, $q=0,1,2$, between the galaxy distribution $\gamma$ and the iSW temperature perturbation $\tau$ we also calculate for the 
first time the mixed trispectra of the analogous form $\bra \tau^q \gamma^{4-q}\ket$. Both types of spectra were consistently derived in tree-level 
perturbation theory in Newtonian gravity. This implies for the bispectra perturbative corrections to second order and for the mixed trispectra contributions from second and third 
order terms. Furthermore, we investigated the time evolution of these individual 3-dimensional source terms, which are in general very diverse. 
For this reason, the time evolution and the configuration dependence of a specific class of spectra, equilateral bispectra and the square trispectra, were studied. 
Finally, the achievable signal-to-noise ratios were derived for measurements cross-correlating \textit{Planck} data and a galaxy sample, as it would be expected from 
a wide angle survey as \textit{Euclid}. 

(i) 
The linear iSW-effect has the time dependence $\mathrm d (D_+/a)/ \mathrm d a$, which makes it sensitive to dark energy but vanishes in SCDM-models with 
$\Omega_\mathrm m \equiv 1$ 
and $D_+(a) = a$. In contrast to this, the non-linear contributions to the iSW signal are sensitive to derivatives of higher powers of $D_+(a)$, namely 
$\mathrm d (D_+^2/a)/ \mathrm d a$ for second order perturbation theory and $\mathrm d (D_+^3/a)/ \mathrm d a$ for third order contributions. For this reason, the 
effect does not vanish in matter-dominated epochs.

(ii) 
The covariances of the measurements were derived in a Gaussian approximation. For the CMB observation the intrinsic CMB fluctuations and instrumental noise in form 
of the pixel noise and a Gaussian beam were considered as noise sources. A Poissonian noise term was added to the galaxy distribution signal. For simplicity the 
fluctuations of the dark matter density and galaxy number density were related to each other by a constant linear biasing model.

(iii) 
In the mixed bispectra and trispectra the configuration and scale dependence represent the different correlation lengths of the gravitational potential and the 
density field. Since the specific perturbative corrections dominate on different scales, the mixed spectra change their sign at certain values of $\ell$. 
In case of the bispectra one can observe the transition from linear domination to non-linear domination move to larger and larger scales with increasing number of 
included iSW source fields $q$.

(iv) 
We derived the cumulative signal-to-noise ratios $\Sigma_q^{(3)}$ for the measurements of mixed bispectra $\bra \tau^q \gamma^{3-q}\ket$, 
and  $\Sigma_q^{(4)}$ for the mixed trispectra of the form
$\bra \tau^q \gamma^{4-q}\ket$, with a Gaussian approximation to the covariance. The integration were performed numerically
using Monte Carlo integration techniques from the multidimensional numerical integration library CUBA (Hahn 2005). 
For both spectra the initial CMB fluctuations are the most important noise source, which makes it difficult to observe the signals. We assumed 
a cross-correlation of \textit{Planck} data with a \textit{Euclid}-like galaxy sample starting from angular scales of $\ell_\mathrm{min}=10$ up to a resolution of 
$\ell_\mathrm{max}=10^3$. The only spectra reaching the order of magnitude of the noise level are the bispectra and trispectra in the configuration 
$\bra\tau\gamma^{n-1}\ket$. We found the numerical signal-to-noise ratios of $\Sigma_1^{(3)}=0.828$ for the bispectrum and $\Sigma_1^{(4)}=0.19$ for the 
trispectrum and conclude, that non-Gaussian signatures of the iSW-effect are too weak to be detected. At the same time, 
these small signal-to-noise ratios suggest that non-Gaussianities in the CMB generated by the iSW-effect are small enough so that 
they do not interfere with the estimation of the inflationary non-Gaussianity parameter $f_\mathrm{NL}$ from the bispectrum 
$\bra\tau^3\ket$ and of the two parameters $g_\mathrm{NL}$ and $\tau_\mathrm{NL}$ from the trispectrum $\bra\tau^4\ket$.
\section*{ACKNOWLEDGEMENTS}
We would like to thank Matthias Bartelmann for useful discussions and ideas. Our work was supported by the German Research Foundation (DFG) within the framework of the Priority Programme 1177 and the excellence 
initiative through the Heidelberg Graduate School of Fundamental Physics. 

\bibliography{bibtex/aamnem,bibtex/references}
\bibliographystyle{mn2e}

\appendix
\section{Analytical details of signal-to-noise ratios}
\label{ap1}
The squared signal-to-noise ratio $\Sigma^2$ is given by the $\chi^2$ between a detection and its zero hypothesis. In the course of its calculation, one has to 
ensure that no redundant information is taken into account. We present the calculation for the bispectra only, since it follows the same 
argumentation in the case of the trispectra. Neglecting redundancy due to any symmetries all mixed and pure bispectra would account for a $\chi^2$-
contribution of
\ba
\chi^2 &=& \frac {f_\mathrm{sky}}{\pi}\frac 1{(2\pi)^2} 
      \int \mathrm d^2 \bl_{1,2,3}\, \mathrm d^2 \bl_{1^\prime ,2^\prime, 3^\prime}\,\, \dirac(\bl_1+\bl_2+\bl_3)\nn \\
      ~&~&
       B_{i_1i_2i_3}^{\,\ell_1,\ell_2,\ell_3} 
  \,\mathrm{Cov}^{-1}\left[\tilde B_{i_1i_2i_3}^{\,\ell_1,\ell_2,\ell_3}, \tilde B_{i_1^\prime i_2^\prime i_3^\prime}^{\,\ell_1^\prime,\ell_2^\prime,\ell_3^\prime}\right] \,
      \tilde B_{i_1^\prime i_2^\prime i_3^\prime}^{\,\ell_1^\prime,\ell_2^\prime,\ell_3^\prime}\,,
\ea
where also the sum over all field indices is implied.
However, the integrand is symmetric in any simultaneous pairwise permutation of $(\bl_n,i_n)$ with $(\bl_m,i_m)$ and likewise of $(\bl_n^\prime,i_n^\prime)$ with $(\bl_m^\prime,i_m^\prime)$. 
This type of redundancy can be avoided by constraining the integration volumes to $\ell_1<\ell_2<\ell_3$ and $\ell_1^\prime<\ell_2^\prime<\ell_3^\prime$. Furthermore, 
the sum over field indices may lead to more redundancy, which we encode at this point into a multiplicity factor $s_{i_1i_2i_3}^{i_1^\prime i_2^\prime i_3^\prime}$. 
Now, the signal-to-noise ratio can be written as
\begin{multline}
\label{ap:s2n}
\left(\Sigma^{(3)}\right)^2 = \frac {f_\mathrm{sky}}{4\pi^3} 
      \int \limits_{\ell_1<\ell_2<\ell_3}\mathrm d^2 \bl_{1,2,3}\,\int \limits_{\ell_1^\prime<\ell_2^\prime<\ell_3^\prime} \mathrm d^2 \bl_{1^\prime ,2^\prime, 3^\prime}\,
\, \dirac(\bl_1+\bl_2+\bl_3)\\
       \times B_{i_1i_2i_3}^{\,\ell_1,\ell_2,\ell_3} 
  \,\mathrm{Cov}^{-1}\left[\tilde B_{i_1i_2i_3}^{\,\ell_1,\ell_2,\ell_3}, \tilde B_{i_1^\prime i_2^\prime i_3^\prime}^{\,\ell_1^\prime,\ell_2^\prime,\ell_3^\prime}\right] \,
      \tilde B_{i_1^\prime i_2^\prime i_3^\prime}^{\,\ell_1^\prime,\ell_2^\prime,\ell_3^\prime}\,\left(s_{i_1i_2i_3}^{i_1^\prime i_2^\prime i_3^\prime}\right)^{-1}\,,
\end{multline}
In this subspace the covariance matrix can be inverted, as it was shown in Section~\ref{sect_cov}. Substituting eqn.~(\ref{cov_inv_bi}) into eqn.~(\ref{ap:s2n}), we find
\begin{multline}
\label{ap:s2n1}
\left(\Sigma^{(3)}\right)^2 = \frac {f_\mathrm{sky}}{4\pi^3} \,\left(s_{i_1i_2i_3}^{i_1^\prime i_2^\prime i_3^\prime}\right)^{-1}
      \int \limits_{\ell_1<\ell_2<\ell_3}\mathrm d^2 \bl_{1}\,\mathrm d^2 \bl_{2}\, \\
       \times B_{i_1i_2i_3}^{\,\ell_1,\ell_2,\ell_3} 
  \,\frac{\tilde C_{i_1^{~} i_1^\prime}^{\ast\,\ell_1}\tilde C_{i_2^{~} i_2^\prime}^{\ast\,\ell_2}\tilde C_{i_3^{~} i_3^\prime}^{\ast\,\ell_3} }
{\det C^{\,\ell_1}\det C^{\,\ell_2}\det C^{\,\ell_3}}
      \tilde B_{i_1^\prime i_2^\prime i_3^\prime}^{\,\ell_1,\ell_2,\ell_3}\,,
\end{multline}
where from now on $\bl_3=-\bl_1-\bl_2$ is implied, if $\bl_3$ is not integrated over. 
If one is now interested in the signal-to-noise ratio of particular field mixture, i.e. data with a fixed field configuration $q=i_1+i_2+i_3$, one can further simplify the 
expression. For pure galaxy contributions, $q=q^\prime=0$, we can neglect the cross-correlation , i.e. $C_{01}=0$, and no redundancy due to field summation occurs, 
$s_{000}^{000}=1$. One obtains the well-known case \citep{Hu2001}
\be
\label{ap:s2n_q=0}
\left(\Sigma_0^{(3)}\right)^2 = \frac {f_\mathrm{sky}}{4\pi^3} \,
      \int \mathrm d^2 \bl_{1}\,\mathrm d^2 \bl_{2}\,\frac{\left( B_{000}^{\,\ell_1,\ell_2,\ell_3} \right)^2}
{6\,\tilde C_{00}^{\,\ell_1}\tilde C_{00}^{\,\ell_2}\tilde C_{00}^{\,\ell_3} }\,,
\ee
where the symmetry in the integrand was used to obtain an integration over full $\bl$-space in combination with the factor $1/6$. 
If all mixed spectra with one iSW field are taken into account, i.e. $q=q^\prime=1$, one obtains $9$ different contributions due to the field index summation.
Three contributions are quadratic in identical bispectra, $(i_1,i_2,i_3) = (i_1^\prime,i_2^\prime,i_3^\prime)$, and have multiplicity one. The remaining mixed contributions 
have multiplicity 2, since the integrand in eqn.~(\ref{ap:s2n1}) is symmetric under exchange of the primed and unprimed index sets 
$(i_1,i_2,i_3)$ and $(i_1^\prime,i_2^\prime,i_3^\prime)$ as a whole. Therefore the multiplicities are
\ba
1&=&s_{001}^{001} = s_{010}^{010} = s_{100}^{100}\nn \\
2&=& s_{001}^{010} = s_{001}^{100} = s_{010}^{100} = s_{010}^{001} = s_{100}^{001} = s_{100}^{010}\,.
\ea
If one uses
\ba
B_{010}^{\,\ell_1,\ell_2,\ell_3} &=& B_{001}^{\,\ell_1,\ell_3,\ell_2}\nn \\
B_{100}^{\,\ell_1,\ell_2,\ell_3} &=& B_{001}^{\,\ell_3,\ell_2,\ell_1}
\ea
in combination with
\ba
B_{001}^{\,\ell_1,\ell_2,\ell_3} &=&\frac 12 \left(B_{001}^{\,\ell_1,\ell_2,\ell_3}+B_{001}^{\,\ell_2,\ell_1,\ell_3}\right)\,,
\ea
one can combine the quadratic terms to one, which is integrated over the full $\bl_{1,2,3}$-volume. This can be done for the mixed terms in analogy and one is left 
with the following expression for the signal-to-noise ratio,
\begin{multline}
\left( \Sigma_1^{(3)} \right)^2 = \frac {f_\mathrm{sky}}{4\pi^3}
      \int \mathrm d^2 \bl_{1}\, \mathrm d^2 \bl_{2}\,\,\left(2\,\det C^{\,\ell_1}\det C^{\,\ell_2}\det C^{\,\ell_3}\right)^{-1}\\
      \left[\left(B_{001}^{\,\ell_1,\ell_2,\ell_3}\right)^2\tilde C_{11}^{\,\ell_1}\tilde C_{11}^{\,\ell_2}\tilde C_{00}^{\,\ell_3}+B_{001}^{\,\ell_1,\ell_2,\ell_3}
\tilde C_{11}^{\,\ell_1}\tilde C_{01}^{\,\ell_2}\tilde C_{01}^{\,\ell_3} B_{010}^{\,\ell_1,\ell_2,\ell_3}\right]\,.
\end{multline}
Following the analog path of argumentation one finds the signal-to-noise expressions for the trispectra to be
\ba
\left( \Sigma_0^{(4)} \right)^2 &=& \frac {f_\mathrm{sky}}{8\pi^4} 
      \int \mathrm d^2 \bl_{1}\, \mathrm d^2 \bl_{2}\, \mathrm d^2 \bl_{3}\,\,
       \frac{\left(T_{0000}^{\,\ell_1,\ell_2,\ell_3,\ell_4}\right)^2}{24\,\tilde C_{00}^{\,\ell_1}\tilde C_{00}^{\,\ell_2}\tilde C_{00}^{\,\ell_3}\tilde C_{00}^{\,\ell_4}}\nn \\
\left( \Sigma_1^{(4)} \right)^2 &=& \frac {f_\mathrm{sky}}{8\pi^4} 
      \int \mathrm d^2 \bl_{1}\, \mathrm d^2 \bl_{2}\, \mathrm d^2 \bl_{3}\,\,\left(\det C^{\,\ell_1}\dots\det C^{\,\ell_4}\right)^{-1}\nn\\
       ~&~& \left[\,\,\,\frac{1}{6}\,\left(T_{0001}^{\,\ell_1,\ell_2,\ell_3,\ell_4}\right)^2 \tilde C_{11}^{\,\ell_1}\tilde C_{11}^{\,\ell_2}\tilde C_{11}^{\,\ell_3}\tilde C_{00}^{\,\ell_4}\right. \nn \\
       ~&~& +\,\frac{1}{4}\,\left.T_{0001}^{\,\ell_1,\ell_2,\ell_3,\ell_4}\tilde C_{11}^{\,\ell_1}\tilde C_{11}^{\,\ell_2}\tilde C_{01}^{\,\ell_3}\tilde C_{01}^{\,\ell_4}T_{0010}^{\,\ell_1,\ell_2,\ell_3,\ell_4}\,\right]\,.
\ea
Also the expressions for higher values of $q$ can now be deduced with the same techniques.

\bsp

\label{lastpage}

\end{document}